\documentclass[12pt,draftclsnofoot,onecolumn]{IEEEtran}

\usepackage{amssymb}
\usepackage{amsmath}
\usepackage{amsfonts}
\usepackage{graphicx}
\usepackage{subfigure}
\usepackage{cite}
\usepackage{enumerate}
\usepackage{url}
\usepackage[noend]{algpseudocode}
\usepackage{algorithmicx,algorithm}
\usepackage{mathtools}
\usepackage{exscale}
\usepackage{relsize}
\allowdisplaybreaks[4]

\begin{document}

\title{\LARGE{Ultra-dense LEO: Integrating Terrestrial-Satellite Networks into 5G and Beyond for Data Offloading}}
\author{
\IEEEauthorblockN{
\normalsize{Boya Di}, \emph{Student Member, IEEE},
\normalsize{Hongliang Zhang}, \emph{Student Member, IEEE},
\normalsize{Lingyang~Song}, \emph{Senior Member, IEEE},
\normalsize{Yonghui Li}, \emph{Senior Member, IEEE},
\normalsize{Geoffrey Ye Li}, \emph{Fellow, IEEE}\\}
\thanks{B.~Di, H.~Zhang, and L.~Song are with State Key Laboratory of Advanced Optical Communication Systems and Networks, School of Electronics Engineering and Computer Science, Peking University, Beijing, China. E-mail: \{boya.di, hongliang.zhang, lingyang.song\}@pku.edu.cn.}
\thanks{Y.~Li is with School of Electrical and Information Engineering, the University of Sydney, Australia. E-mail: yonghui.li@sydney.edu.au.}
\thanks{G.~Y.~Li is with School of Electrical and Computer Engineering, Georgia Institute of Technology, USA. E-mail: liye@ieee.org.}
}
\maketitle

\vspace{-1.6cm}
\begin{abstract}
In this paper, we propose a terrestrial-satellite network (TSN) architecture to integrate the ultra-dense low earth orbit (LEO) networks and the terrestrial networks to achieve efficient data offloading. In TSN, each ground user can access the network over C-band via a macro cell, a traditional small cell, or a LEO-backhauled small cell (LSC). Each LSC is then scheduled to upload the received data via multiple satellites over Ka-band. We aim to maximize the sum data rate and the number of accessed users while satisfying the varying backhaul capacity constraints jointly determined by the LEO satellite based backhaul links. The optimization problem is then decomposed into two closely connected subproblems and solved by our proposed matching algorithms. Simulation results show that the integrated network significantly outperforms the non-integrated ones in terms of the sum data rate. The influence of the traffic load and LEO constellation on the system performance is also discussed.
\end{abstract}
\vspace{-0.5cm}

\newpage
\section{Introduction}
In the past decades, the rapid development of cellular communications~\cite{li2017radio} has triggered users' increasing demand for high-data-rate applications~\cite{song2010evolved,wu2016delay}, which in turn raise stringent requirements on achieving massive connectivity and high capacity in the 5G systems\footnote{Typical high-data-rate applications include video streaming~\cite{wu2016delay,wu2016energy}, environment sensing, etc.}. To meet such demands, the heterogeneous network~\cite{zhang2015poster,wu2016energy} has been considered as a well accepted architecture due to its flexible deployment and effective data offloading. However, there still remain several unsolved issues. For example, the network coverage cannot be guaranteed due to the sparse resources and intractable access point deployment~\cite{5G}. Moreover, it is usually not practical to construct fiber-equipped or stable wireless backhaul links for every small cell~\cite{ge20145g} due to the complex environments. Therefore, the limited backhaul capacity of small cells may lead to a degraded offloading performance.


Fortunately, recent advances in low earth orbit (LEO) satellite networks over the high-frequency band have provided an alternative solution for coverage extension and backhaul connectivity. Driven by SpaceX~\cite{SpaceX} and OneWeb~\cite{OneWeb}, the ongoing LEO constellation projects plan to launch thousands of LEO satellites over the earth, aiming to deploy an ultra-dense (UD) constellation and cooperate with the traditional operators to support seamless and high-capacity communication services. With provisioning of the feasibility, the projects perform the pipeline production of small satellites to lower the manufacturing cost~\cite{oneweb-cost}. Instead of the traditional small cell base station (SBS), a dedicated terrestrial-satellite terminal (TST) equipped with steerable antennas acts as the access point in this case. Easy to install on the roof or eNodeB owing to its miniaturized antennas, each TST supports both the high-quality TST-satellite backhaul links over Ka-band and the user-TST links over C-band, enabling a terrestrial small cell coverage for the users. Compared to the traditional networks, the LEO network provides a great number of users with a high-capacity backhaul, vast coverage, and more flexible access technique, which is less dependent of real environments.



To fully exploit the LEO constellation technique, we propose a terrestrial-satellite architecture to integrate the LEO satellite networks and the terrestrial networks for traffic offloading in this paper. Each user is scheduled to upload its data via the macro cell, the traditional small cell (TSC), or the LEO-based small cell (LSC). Each SBS and TST then upload the collected data to the core network via the traditional backhaul and LEO-based backhaul, respectively. Benefited from the UD-LEO constellation, we assume that each TST is allowed to connect to multiple satellites simultaneously~\cite{SpaceX,SANSA}, thereby improving the resilience to the frequent handover of satellites~\cite{papapetrou2004satellite}. Deployed by the same operator, all cells share the same C-band frequency resources for terrestrial communications. Multiple TST-satellite backhaul links over Ka-band are scheduled for each LSC. The network aims to maximize the sum rate of all users and accommodate as many users as possible. Therefore, the user association and resource allocation of multiple cells should be optimized subject to the backhaul capacity constraint of each cell. At the same time, the backhaul capacity of each LSC also needs to be maximized via the satellite selection and resource allocation.

Various techniques have been considered for traffic offloading in the heterogeneous networks, such as satellite access~\cite{vazquez2016hybrid,kuang2017radio}, hybrid satellite-terrestrial relaying~\cite{bhatnagar2015performance,arti2016two,bhatnagar2013performance}, device-to-device (D2D) multi-cell interference management~\cite{zhou2016energy,zhou2017energy,ng2012energy}, and cloud radio access. In~\cite{vazquez2016hybrid}, the terrestrial-satellite backhaul network shares the Ka-band and the terrestrial wireless links deploy the hybrid analog-digital transmit beamforming to mitigate interference. In~\cite{kuang2017radio}, a single satellite covers a large area and the integrated terrestrial-satellite networks perform the cloud-based resource allocation. In~\cite{bhatnagar2015performance,arti2016two,bhatnagar2013performance}, a hybrid terrestrial-satellite network sharing the same spectrum has been considered where either the satellite serves as a relay for one terrestrial source-destination pair or a terrestrial node relayes for the satellite -- user pair. The average symbol error rate has been derived and the analytical diversity order has been obtained. In~\cite{zhou2016energy,zhou2017energy}, the D2D underlaying dense network has been considered and a distributed algorithm has been proposed to cope with the strong intra-cell and inter-cell interference. In~\cite{ng2012energy}, energy-efficient user association and resource allocation has been investigated in a multi-cell downlink network under the constraints of the backhaul capacity.

Most of the existing works~\cite{vazquez2016hybrid,kuang2017radio,bhatnagar2015performance,arti2016two,bhatnagar2013performance,zhou2016energy,zhou2017energy,ng2012energy} have assumed either ideal or fixed backhaul capacity. Differently, in this paper we aim to improve the terrestrial-satellite system performance by considering the influence of dynamically varying backhaul capacity determined by the satellite selection and Ka-band resource allocation. Moreover, the multi-connectivity also provides new dimension for backhaul capacity optimization, but at the same time new challenges have thus been posed on the traffic offloading strategy design. On the one hand, the LEO-based network induces long propagation delay even though it has high-capacity backhaul links. Combined with the backhaul transmission delay, the overall delay will influence the user scheduling strategy. Therefore, scheduling and resource allocation of both terrestrial and satellite networks are coupled. On the other hand, due to the angle-sensitivity and the multi-connectivity, the UD-LEO satellite networks may suffer from intractable inter-satellite co-channel interference. The influence of angular separation between two TST-satellite links should be carefully considered to improve the LEO-based backhaul capacity.

The main contribution of this paper can be summarized as below:
\begin{itemize}
\item We propose a scheme for data offloading in the heterogenous networks by utilizing the high-capacity LEO-based backhaul. To perform the user scheduling in such a network, a joint optimization problem is formulated to maximize the sum rate and number of accessed users subject to the backhaul capacity constraints.
\item The formulated optimization problem is decomposed into two subproblems, i.e., the maximization of the sum rate and number of accessed users in the terrestrial networks and the total backhaul capacity maximization in the satellite networks, closely connected by the iteratively varying Lagrangian multiplexers. We then convert these two subproblems into two matching problems with externalities. For the first subproblem, a low-complexity modified Gale-Shapely matching algorithm~\cite{roth1992two} is proposed. For the second one, we develop a novel swap matching algorithm with power control and gradient-based pruning procedures, in which the angle-sensitive nature is captured.
\item Based on computer simulation results, the performance of our proposed UD-LEO based integrated terrestrial-satellite (LITS) scheme significantly outperforms the traditional non-integrated networks. We also find that the traffic load and different LEO satellite constellations will influence the user scheduling strategy.
\end{itemize}

The rest of this paper is organized as follows. In Section \uppercase\expandafter{\romannumeral2}, we describe the model of integrated UD-LEO based terrestrial-satellite networks. In Section \uppercase\expandafter{\romannumeral3}, we formulate a joint optimization problem for the terrestrial-satellite networks and propose a framework to iteratively solve two decoupled subproblems from Lagrangian dual decomposition. In Sections \uppercase\expandafter{\romannumeral4} and \uppercase\expandafter{\romannumeral5}, we convert these two subproblems into two different matching problems and solve them. Simulation results are presented in Section \uppercase\expandafter{\romannumeral6}, and finally, we conclude the paper in Section \uppercase\expandafter{\romannumeral7}.


\section{System Model}
In this section, we first introduce the UD-LEO based integrated terrestrial-satellite network in which the users can access the network via a macro BS or the TSCs or the LSCs. We then provide the transmission models of both the terrestrial and satellite communications. For convenience, we summarize all non-standard abbreviations in Table~\ref{abbreviation}.

\begin{table}
	\begin{center}
		\caption{Major Non-standard Abbreviations}
		\label{abbreviation}
		\begin{tabular}{|l|l|}
			\hline
			\bf{Abbreviation} & \bf{Paraphrase}\\
			\hline LITS & LEO-based integrated terrestrial-satellite scheme\\
			\hline LBCO & LEO-based backhaul capacity optimization\\
			\hline LSC & LEO-based small cell\\
			\hline PC & Power control\\
			\hline NITS & Non-integrated terrestrial-satellite networks\\
			\hline SBS & Small cell base station\\
			\hline SMPC & Swap matching algorithm with power control\\
			\hline TSC  & Traditional small cell\\
			\hline TST & Terrestrial-satellite terminal\\
			\hline TTH & Traditional terrestrial heterogeneous networks\\
			\hline TTO & Terrestrial traffic offloading\\
			\hline TUASA & Terrestrial user association and subchannel allocation\\
			\hline UD & Ultra-dense\\
			\hline
		\end{tabular}
	\end{center}
\end{table}

\vspace{-0.3cm}
\subsection{Scenario Description}
\begin{figure}[!t]
	\centering
	\includegraphics[width=6.3in]{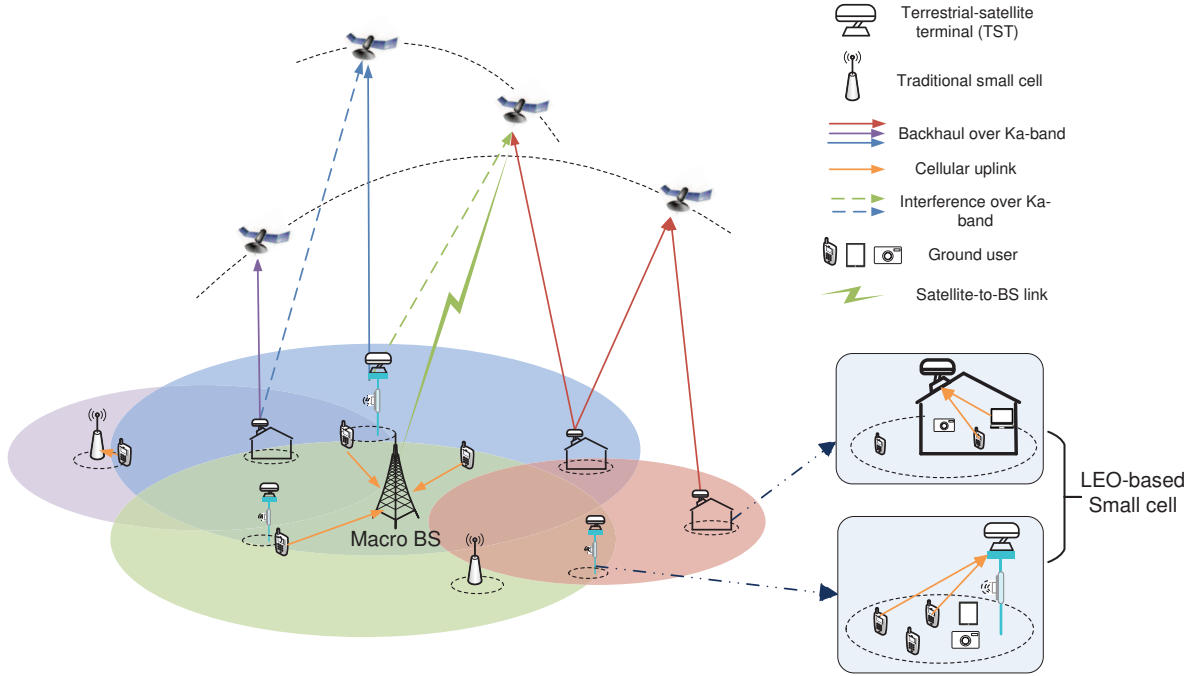}
	\caption{System model of the ultra-dense LEO-based integrated terrestrial-satellite network} \label{system_model}
\end{figure}

Consider an UD-LEO based integrated terrestrial-satellite network as shown in Fig.~\ref{system_model}, where one macro BS and a large number of small cells are deployed to serve the uplink ground users\footnote{The proposed architecture is independent of the satellite altitude and constellation, and thus, can be directly extended to the MEO and GEO satellite systems. The data routing and signaling control problems need to be considered for the multi-layer system, which is not the focus of this paper. Readers may find an initial work focusing on contact graph based routing in~\cite{shi2017traffic}.}. Each small cell assists the macro cell to offload the traffic and is connected to the core networks via either wired or wireless backhauls. Therefore, each user, such as a mobile device or a sensor, can access the network via one of the following three cells: 1) the macro cell with large backhaul capacity supported by fiber links from the macro BS directly to the core network; 2) the TSC with very limited backhaul capacity connected to the core network via multi-hop wired or wireless backhaul links; 3) the LSC with large backhual capacity supported by the Ka-band transmission. For LSC backhaul, each TST uploads the user's data to the LEO satellites, and then each satellite forwards it to either an earth gateway station (connected to the core network) or a TST-equipped macro BS.

The ultra-dense LEO topology ensures that multiple satellites fly over the area of interest at each time slot, providing a seamless coverage for the mobile users. Equipped with multiple independent antenna apertures~\cite{SpaceX,OneWeb}, each TST can connect to multiple satellites simultaneously, which further improves the backhaul capacity of the LSCs. Based on the current development of satellite technique~\cite{OneWeb}, we assume that each satellite serves as a remote radio head of the BS with no on-board processing capacity. Therefore, the access control is located at the macro BS equipped with a TST~\cite{3GPP2017}. For inter-cell interference management, we assume that the satellite operator and the traditional terrestrial operator cooperate to serve the users in a centralized manner.


\vspace{-0.3cm}
\subsection{Transmission Model for Terrestrial Communications}

Denote the set of cells as ${\cal{M}} = \left\{ {{\rm{0}}}, {\rm{1}}, \cdots, M \right\}$ in which $m = 0$ represents the macro BS, $1 \le m \le M'$ represents the TSCs, and $M'+1 \le m \le M$ indicates the LSCs. To describe the relationship between the positions of $J$ users and the coverage of each cell, a binary coverage matrix $\emph{\textbf{A}}$ of size $(M+1) \times J$ is introduced where $a_{m,j} = 1$ indicates that user $j$ lies in within the coverage of cell $m$, and $a_{m,j} = 0$ otherwise.

For frequency re-use, all cells share the same frequency resource pool, which can be divided into $K$ subchannels. To better depict the user association and subchannel allocation, we introduce a binary matrix $\emph{\textbf{X}}$ of size $(M+1) \times J \times K$ in which $x_{m,j,k} = 1$ indicates that user $j$ is served by BS $m$ over subchannel $k$, and $x_{m,j,k} = 0$ otherwise. The received signal of BS $m$ sent by user $j$ over subchannel $k$ is then given by
\vspace{-0.3cm}
\begin{equation} \label{signal_model}
{y_{m,j,k}} = \sqrt {{p_u}} h_{m,j,k}^B{x_{m,j,k}}{s_j} + \underbrace{\sum\limits_{m' \ne m} {\sum\limits_{j' \ne j} {\sqrt {{p_u}} h_{m,j',k}^B{x_{m',j',k}}{s_{j'}}} }}_{\mbox{inter-cell co-channel interference}}  + {N_m},
\end{equation}
where $p_u$ is the transmit powers of each user, $s_j$ (or $s_{j'}$) is the transmitted signal of user $j$ (or user $j'$) with unit energy, and the corresponding channel coefficient is $h_{m,j,k}^B$ (or $h_{m,j',k}^B$). Specifically, we denote $h_{m,j,k}^B = g_{m,j,k} \cdot \beta_{m,j} \cdot {\left( {d_{m,j}} \right)^{ - \alpha }}$, where $g_{m,j,k} \sim {\cal{CN}}{(0,1)}$ is a complex Gaussian variable representing Rayleigh fading, $\beta_{m,j}$ follows log-normal distribution representing shadowing fading, $d_{m,j}$ is the distance between users $j$ and $m$, and $\alpha$ represents the pathloss exponent. The additive white Gaussian noise~(AWGN) at BS $m$ is denoted by ${N_{m}} \sim \mathcal{CN}\left( {0,{\sigma}^2} \right)$, and ${\sigma}^2$ is the noise variance.

The achievable rate of each user $j$ served by BS $m$ over subchannel $k$ can be expressed by
\begin{equation} \label{single_rate}
{R_{m,j,k}^B} = {\log _2}\left( {1 + \frac{{{p_u}{{\left| {h_{m,j,k}^B} \right|}^2}}}{{{\sigma ^2} + \sum\limits_{j' \ne j} {\sum\limits_{m' \ne m} {{x_{m',j',k}}{p_u}{{\left| {h_{m,j',k}^B} \right|}^2}} } }}} \right),
\end{equation}
and thus, the data rate of each cell $m$ can be obtained by
\vspace{-0.1cm}
\begin{equation}\label{single_cell_rate}
R_m^B = \sum\nolimits_{j = 1}^J {\sum\nolimits_{k = 1}^K {{x_{m,j,k}}{R_{m,j,k}^B}} }.
\end{equation}
\vspace{-0.1cm}
In practice, the data rate of each cell is required to be no larger than the backhaul capacity of this cell, which also influences the traffic offloading scheme of the system. For convenience, we denote the backhaul capacity of each cell $m$ as $C_m$, which is fixed when $0 \le m \le M'$.
\vspace{-0.4cm}
\subsection{Transmission model for LEO-based Backhaul}
Instead of a fixed value, the backhaul capacity of each LSC is related to the transmission model of the LEO-based backhaul over Ka-band. Note that the satellite-to-ground links usually occupy a wider bandwidth than the TST-to-satellite links and the transmit power of satellites is usually larger than that of the TSTs~\cite{SpaceX}. The LEO backhaul capacity is thus constrained by the TST-to-satellite links, as illustrated below.

\begin{figure}
	\centering
	\subfigure[]{
		\label{antenna} 
		\includegraphics[width=2.8in]{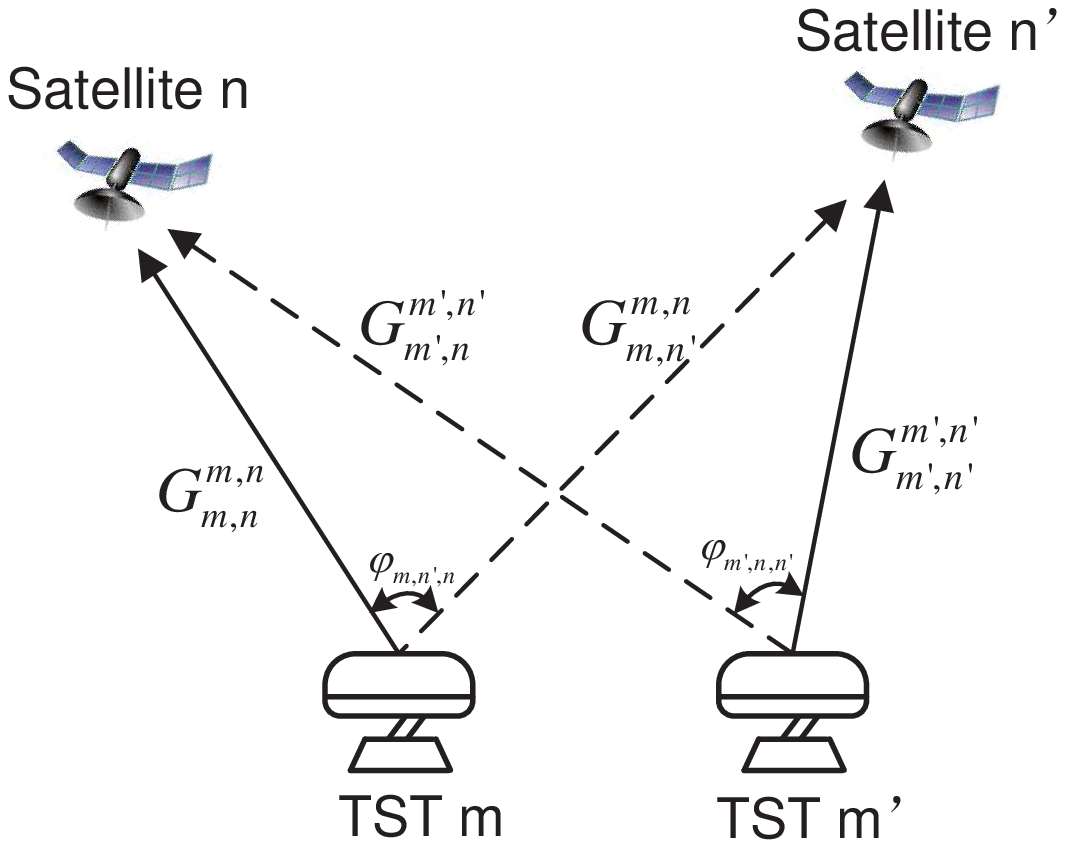}}
	\hspace{-0.2in}
	\subfigure[]{
		\label{angle} 
		\includegraphics[width=3in]{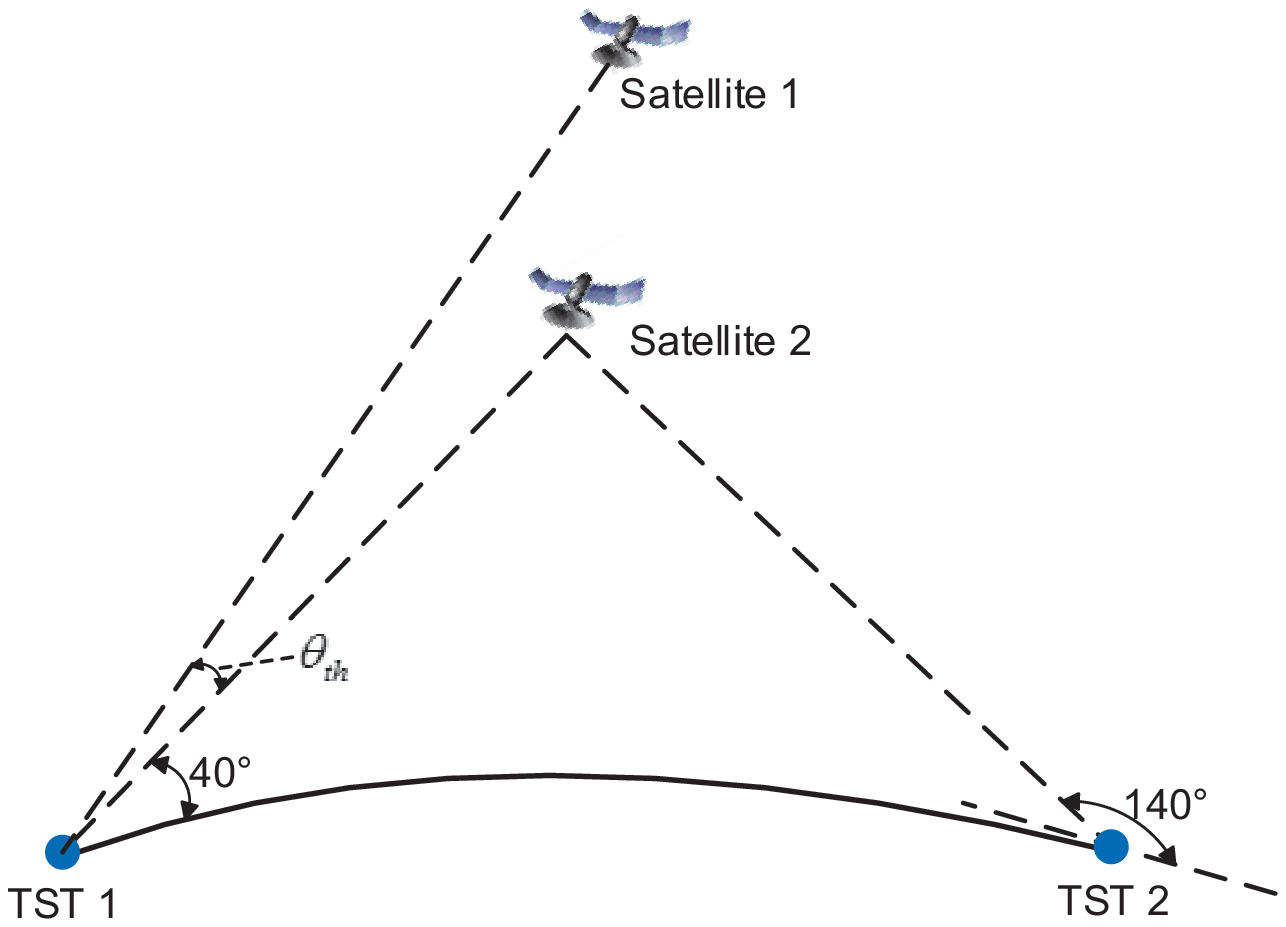}}
	\vspace{-0.1cm}
	\caption{(a) Illustration of each TST's antenna gain; (b) Illustration of each satellite's service range and the angular separation.} \label{doublelast} 
\end{figure}

We assume that there are $N$ satellites flying over the area of interest. Due to the pre-planned orbit of each satellite, its altitude, speed, and position information are known to all cells in each time slot. For convenience, we adopt a qausi-static method to split a time period into multiple time slots\footnote{To save the signaling cost, we adopt the semi-persistent scheduling where the LEO backhaul links are updated every few time slots.} during each of which the position of a satellite is unchanged. We divide the available bandwidth over the Ka-band spectrum as a set of subchannels ${{\mathcal{Q}}} = \left\{ {1,...,Q} \right\}$. A binary matrix depicting the TST-LEO association and subchannel allocation is defined as $\emph{\textbf{B}}$ where the element $b_{m,n,q} = 1$ indicates that TST $m$ is associated with satellite $n$ over subchannel $q$ and $b_{m,n,q} = 0$ otherwise. The received signal of satellite $n$ sent by TST $m$ over subchannel $q$ can be given by
\begin{equation} \label{signal_LEO}
{y_{m,n,q}} = \sqrt {p_{m,n,q}^TG_{m,n}^{m,n}} h_{m,n,q}^Ts_{m,n}^T{b_{m,n,q}} +\!\! \underbrace{\sum\limits_{m = M' + 1}^M {\sum\limits_{n' \ne n} {\sqrt {p_{m',n',q}^TG_{m',n}^{m',n'}} h_{m',n,q}^Ts_{m',n'}^T{b_{m',n',q}}} } }_{\mbox{inter-satellite co-channel interference}} + {N_n},
\end{equation}
where $p_{m,n,q}^T$ (or $p_{m',n',q}^T$) is the transmit power of TST $m$ to satellite $n$ (or TST $m'$ to satellite $n'$) over subchannel $q$, $s_{m,n}^T$ (or $s_{m',n'}$) is the transmitted signal of TST $m$ to satellite $n$ (or TST $m'$ to satellite $n'$). The channel gain of the TST $m$ -- satellite $n$ link over subchannel $q$ is denoted by $h_{m,n,q}^T$, with both the large-scale fading and the shadowed-Rician fading~\cite{abdi2003new} taken into consideration\footnote{The channel state information (CSI) here can be obtained by adopting the widely utilized training data based CSI estimation techniques~\cite{chaouech2010channel,arti2015imperfect,arti2016channel}. For reference, the p.d.f. of the channel coefficient is given as ${f_{{{\left| h_{m,n,q}^T \right|}^2}}}\left( x \right) = \alpha {e^{ - \beta x}}_1{F_1}\left( {m';1;\delta x} \right)$, where the parameters $\alpha$, $\beta$, $m'$, and $\delta$ can be found in~\cite{bhatnagar2013performance}.}. As shown in Fig.~\ref{antenna}, $G_{m,n}^{m,n}$ is the antenna gain of TST $m$ towards satellite $n$ and $G_{m',n}^{m',n'}$ is the off-axis antenna gain of TST $m'$ towards the direction of satellite $n$ when the target direction of TST $m'$ is towards satellite $n'$. Denote the angular separation between the TST $m'$ -- satellite $n$ link and the TST $m'$ -- satellite $n'$ link as $\varphi_{m',n,n'}$. The item $G_{m',n}^{m',n'}$ is a function of $\varphi_{m',n,n'}$, as shown in~\cite{regulations2004edition} (Attachment III, Appendix 8). AWGN at satellite $n$ is ${N_{n}} \sim \mathcal{CN}\left( {0,{\sigma}^2} \right)$.

The achievable rate of the TST $m$ -- satellite $n$ link over subchannel $k$ can be obtained by
\begin{equation} \label{rate_LEO}
R_{m,n,q}^T = {\log _2}\left( {1 + \frac{{{b_{m,n,q}}p_{m,n,q}^TG_{m,n}^{m,n}{{\left| {h_{m,n,q}^T} \right|}^2}}}{{\sum\nolimits_{n' \ne n} {\sum\nolimits_{m = M' + 1}^M {{b_{m',n',q}}p_{m',n',q}^TG_{m',n}^{m',n'}{{\left| {h_{m',n,q}^T} \right|}^2}} }  + {\sigma ^2}}}} \right).
\end{equation}
Note that each TST-satellite link suffers the propagation delay due to the long distance compared to the terrestrial communications. Therefore, given the traffic load of each TST, i.e., $L_m$, the equivalent backhaul capacity ${C_m}$ can be given by
\begin{equation} \label{backhaul_capacity}
{C_m} = \sum\limits_{n = 1}^N {\frac{{\rm{1}}}{{{\rm{1/}}{C_{m,n}} + {T_{trip,n}}/{L_{m,n}}}}},
\end{equation}
where ${C_{m,n}} = \sum\nolimits_{q = 1}^Q {R_{m,n,q}^T}$ is the capacity of the TST $m$ -- satellite $n$ link, $L_{m,n}$ is the traffic load over the TST $m$ -- satellite $n$ link, and $T_{trip,n}$ is the propagation delay calculated by $2H_n/c$ with $H_n$ being the altitude of satellite $n$ and $c$ being the speed of light. The values of $L_{m,n}$ can be obtained by solving the following equations:
\begin{equation} \label{equation_set}
\begin{split}
& L_{m,n} / C_{m,n} + T_{trip,n}  = L_{m,n'} / C_{m,n'} + T_{trip,n'}, \forall n,n' \in {\cal{N}}_{m}\\
& \sum\nolimits_{n \in {{\cal{N}}_m}} {{L_{m,n}}}   = {L_m},
\end{split}
\end{equation}
where ${\cal{N}}_m = \{1\le n\le N|b_{m,n,q} = 1, \forall q\}$ is the set of associated satellites for TST $m$. In practice, the traffic load $L_m$ is set as $\min \left\{ {R_m^B,\sum\nolimits_{j \in {{\cal{J}}_m}} {{D_j}} } \right\}$, where ${\cal{J}}_m = \{1\le j\le J|x_{m,j,k} = 1, \forall k\}$ is the set of users associating to cell $m$ and $D_j$ is the amount of data generated by each user $j$. For convenience, we denote the total backhaul capacity without considering the propagation delay as ${\tilde C_m} = \sum\nolimits_{n = 1}^N {{C_{m,n}}}$.

\section{Problem Formulation and Decomposition}
In this section, we aim to maximize the sum rate and number of accessed users by jointly optimizing the terrestrial data offloading, TST-satellite association, and resource allocation.
\vspace{-0.3cm}
\subsection{Angular constraints for LEO backhaul}
To further depict the characteristics of UD-LEO networks such as angular sensitivity on satellite selection, the angular constraint on the terrestrial-satellite links are illustrated below.
	

As shown in Fig.~\ref{angle}, due to different altitudes of the satellites, the elevation angles of two satellites serving the same TST may be the same. When a TST transmits to such two satellites over the same subchannel, they appear to be ``in line" in this case, thereby leading to transmission failure. To cope with this issue, we define the angular separation as $\theta_{th}$ within which two satellites cannot provide service to the same TST over the same subchannel. Based on the pre-planned satellite orbits, we introduce an angle matrix $\Theta$, where the elevation angle of the TST $m$ -- satellite $n$ link is denoted as $\theta_{m,n}$. The angle constraint can then be presented as
\vspace{-0.2cm}
\begin{equation} \label{angle_constraint}
{b_{m,{n_1},q}}{b_{m,{n_2},q}}{\theta _{th}} \le \left| {{\theta _{m,{n_1}}} - {\theta _{m,{n_2}}}} \right|,\forall 1 \le {n_1},{n_2} \le N,M' + 1 \le m \le M,1 \le q \le Q.
\end{equation}

%
%
\subsection{Problem Formulation}
We consider two performance metrics in this work. First, we aim to maximize the sum rate of all cells over the C-band spectrum subject to the backhaul capacity constraints. Second, to extend the coverage of the traditional terrestrial cellular networks, it is necessary to maximize the number of active users in the network. The problem formulation can be shown as below.
\vspace{-0.1cm}
\begin{align}
&\mathop {\max }\limits_{\left\{ {{x_{m,j,k}},{b_{m,n,q}},p_{m,n}^T} \right\}} \sum\limits_{m = 0}^M {R_m^B} + \mu \sum\limits_{m = 0}^M {\sum\limits_{j = 1}^J {\sum\limits_{k = 1}^K {{x_{m,j,k}}} } } \label{system_optimization}\\
\emph{s.t.}~ &\left(\ref{angle_constraint}\right),\nonumber\\
&{x_{m,j,k}} \le {a_{m,j}},\forall 0 \le m \le M,1 \le j \le J,1 \le k \le K,\tag{\ref{system_optimization}$a$}\\
&\sum\limits_{m = 0}^M {\sum\limits_{k = 1}^K {{x_{m,j,k}}} }  \le 1,\forall 1 \le j \le J,\tag{\ref{system_optimization}$b$}\\
&\sum\limits_{j = 1}^J {{x_{m,j,k}}}  \le 1,\forall 0 \le m \le M,1 \le k \le K,\tag{\ref{system_optimization}$c$}\\
&R_m^B \le {C_m},\forall 0 \le m \le M,\tag{\ref{system_optimization}$d$}\\
&\sum\limits_{m = M' + 1}^M {{b_{m,n,q}}}  \le 1,\forall 1 \le n \le N,1 \le q \le Q, \tag{\ref{system_optimization}$e$}\\
&\sum\limits_{n = 1}^N {\sum\limits_{q = 1}^Q {{b_{m,n,q}}} }  \le {N_r},\forall M'+1 \le m \le M, \tag{\ref{system_optimization}$f$}\\
&\sum\limits_{n = 1}^N {\sum\limits_{q = 1}^Q {p_{m,n,q}^T} }  \le {P^T}, \forall M'+1 \le m \le M, \tag{\ref{system_optimization}$g$}\\
&x_{m,j,k},b_{m,n,q} \in \left\{ {0,1} \right\}, 1 \le j \le J, 1 \le k \le K, 0 \le m \le M, 1 \le n \le N, 1 \le q \le Q,\tag{\ref{system_optimization}$h$}
\end{align}
where $\mu$ is the conversion parameter. Constraint $\left( 9a \right)$ implies that each user $j$ can only access a cell within its reach, i.e., user $j$ is within the range of its subscribed cell. For fairness, we assume that each user can only access one cell and be assigned one subchannel, as shown in constraint $\left( 9b \right)$. As it is typical for the uplink in 3GPP, we consider orthogonal use of frequency resources within each cell, which is guaranteed by constraint $\left( 9c \right)$. The relationship between the data rate of each cell and its backhaul capacity is shown in $\left( 9d \right)$, constructing the coupling between the terrestrial and satellite networks.  Without loss of generality, we assume that each subchannel of satellite $n$ can only be assigned to at most one TST and each TST is associated to at most $N_r$ satellite links simultaneously for backhaul, as presented in constraints $\left( 9e \right)$ and $\left( 9f \right)$. The transmit power constraint is presented in $\left( 9g \right)$ where $P^T$ is the maximum transmit power of each TST for backhaul.
\vspace{-0.2cm}
\subsection{Lagrangian Dual Decomposition}

As shown in problem~$\left( \ref{system_optimization} \right)$, the terrestrial data offloading and the satellite selection are coupled with each other via the backhaul capacity constraint. Therefore, we decompose the original problem into two subproblems connected by the Lagrangian multiplexers.

Denote ${\boldsymbol{\lambda}}  = \left\{ {{{\lambda} _m},0 \le m \le M} \right\}$ as a vector containing the Lagrangian multiplexers associated with constraint $\left( 9d \right)$. The Lagrangian associated with problem $\left( \ref{system_optimization} \right)$ is then defined as
\begin{equation}
{\mathcal{L}}\left( \textbf{\emph{X}},\emph{\textbf{B}},\textbf{\emph{p}}^T,{\boldsymbol{\lambda}}  \right) = \sum\limits_{m = 0}^M {R_m^B}  + \mu \sum\limits_{m = 0}^M {\sum\limits_{j = 1}^J {\sum\limits_{k = 1}^K {{x_{m,j,k}}} } } + \sum\limits_{m = 0}^M {{\lambda _m}\left( {{C_m} - R_m^B} \right)}\nonumber
\end{equation}
\begin{equation}\label{Lagrangian}
= \sum\limits_{m = 0}^M {\left( {1 - {\lambda _m}} \right)R_m^B}  + \mu \sum\limits_{m = 0}^M {\sum\limits_{j = 1}^J {\sum\limits_{k = 1}^K {{x_{m,j,k}}} } } +  \sum\limits_{m = 0}^M {{\lambda _m}{C_m}}.
\end{equation}
Correspondingly, the dual optimum is given by
\vspace{-0.2cm}
\begin{equation} \label{Dual_optimum}
{g^*} = \mathop {\min }\limits_{{\boldsymbol{\lambda}} \succeq 0} g\left( {\boldsymbol{\lambda}}  \right) = \mathop {\min }\limits_{{\boldsymbol{\lambda}} \succeq 0} \mathop {\max }\limits_{\{\emph{\textbf{X}},\emph{\textbf{B}},\emph{\textbf{p}}^T\}} {\mathcal{L}}\left( {\emph{\textbf{X}},\emph{\textbf{B}},\emph{\textbf{p}}^T,{\boldsymbol{\lambda}} } \right).
\end{equation}
For given ${\boldsymbol{\lambda}}$, the first item in $\left( \ref{Lagrangian} \right)$ is only determined by $\emph{\textbf{X}}$ and the second item is only determined by $\emph{\textbf{B}}$ and $\emph{\textbf{p}}^T$. Therefore, we can divide the original problem into two subproblems, i.e., the terrestrial traffic offloading problem
\begin{equation}
\begin{split}
(TTO:) &\mathop {\max }\limits_{\left\{ {{x_{m,j,k}}} \right\}} \sum\limits_{m = 0}^M (1-\lambda_m){R_m^B} + \mu \sum\limits_{m = 0}^M {\sum\limits_{j = 1}^J {\sum\limits_{k = 1}^K {{x_{m,j,k}}} } } \label{problem1}\\
\emph{s.t.}~& \left(9a\right), \left(9b\right),  \left(9c\right),
\end{split}
\end{equation}
and the LEO-based backhual capacity optimization problem
\begin{equation}
\begin{split}
(LBCO:) &\mathop {\max }\limits_{\left\{ {{b_{m,n,q}}}, p_{m,n,q}^T\right\}} \sum\limits_{m = M'+1}^M \lambda_m{C_m} \label{problem2}\\
\emph{s.t.}~& \left(\ref{angle_constraint}\right), \left(9e\right), \left(9f\right),  \left(9g\right),
\end{split}
\end{equation}
where the summation is counted from $M'+1$ since $\tilde C_m$ $\left( 0 \le m \le M \right)$ is fixed.

The optimization process consists of multiple iterations and in each iteration $t$ two steps are performed: i) given ${\boldsymbol{\lambda}}^{\left( t\right)}$, the TTO problem and LBCO problem are solved, respectively; ii) update ${\boldsymbol{\lambda}}$ by $\lambda _m^{\left( {t + 1} \right)} = \lambda _m^{\left( t \right)} - \delta^{\left( t \right)} \left( {C_m^{\left( t \right)} - R_m^{B,\left( t \right)}} \right), \forall 0 \le m \le M$ in which $\delta^{\left( t \right)}$ is a monotonically decreasing exponential function of $t$. The iterations will not stop until $|\delta^{\left( t+1 \right)} - \delta^{\left( t \right)}| < \varepsilon$.

For those optimization problems with continuous variables, the outcome obtained from the Lagrangian dual method always satisfies the constraints~\cite{boyd2004convex}. However, this cannot be guaranteed for our formulated problem since there are only binary variables such that the data rate of each cell $m$ does not vary continuously. Therefore, we adjust the user scheduling strategy after step one if the constraint $\left( 12d \right)$ is violated. For those cells violating $\left( 12d \right)$, the associating users are removed one by one following the order of increasing data rate of each user-BS link until $\left( 12d \right)$ is satisfied.

\vspace{-0.3cm}
\section{Algorithm Design for Terrestrial Data Offloading}
In this section, we formulate the TTO problem in $\left( \ref{problem1} \right)$ as a one-to-one matching with externalities and propose a modified Gale-Shapley algorithm with re-defined preference relations.
\vspace{-0.8cm}
\subsection{Matching Problem Formulation for Terrestrial Traffic Offloading}
Note that the TTO problem is a three-dimensional integer programming problem with a non-convex objective function. Aiming at solving this problem by a low-complexity algorithm, we recognize that the user association and subchannel allocation can be regarded as a multivariate matching process. Specifically, the users, BSs, and the subchannels are three sets of players to be matched with each other to maximize the utility, while the interdependencies exist among the users due to the co-channel interference. This enables us to solve the problem by utilizing the matching theory, as shown below.
\subsubsection{Definitions} \label{TTO_definition}
Consider the set of users, BSs, and the subchannels as ${\cal{J}}$, ${\cal{M}}$ and ${\cal{K}}$, which are disjoint with each other. By associating each BS with each subchannel, we construct a new set ${\cal{S}} = {\cal{M}} \times {\cal{K}}$ of size $(M+1) \times K$, where each BS-subchannel unit can be represented by $\left( m, k \right)$. A \emph{matching} $\Psi$ is defined as a mapping from the set $\cal{J} \cup {\cal{S}} \cup \varnothing$ into itself such that for each user $j$ and each BS-subchannel unit $\left( m, k \right)$, we have $\Psi \left( j \right) = \left( {m,k} \right)$ if and only if $\Psi \left( {m,k} \right) = j$. In other words, if $\Psi \left( j \right) = \left( {m,k} \right)$, then user $j$ is associated with BS $m$ over subchannel $k$. Such a one-to-one matching naturally satisfies constraints $\left( 9b \right)$ and $\left( 9c \right)$.


To construct a matching $\Psi$, each BS-subchannel unit selects a user from ${\cal{J}} \cup \varnothing$ to match with such that the weighted sum rate in $\left(\ref{problem1}\right)$ can be maximized. For convenience, when $\Psi \left( j \right) = \left( {m,k} \right)$, we denote the utility of the BS-subchannel unit $\left( m, k \right)$ and user $j$ as $\left( 1 - \lambda_m\right) R_{m,j,k}^B$.
\subsubsection{Preference Relation} \label{TTO_preference}
Due to the inter-cell co-channel interference, there exist externalities~\cite{pycia2015matching} in this matching problem. That is, the utility of $\left( m, k \right)$ is influenced by other users matched with $\left( m', k \right)$. Therefore, each BS-subchannel unit actually has preference over the matching pairs due to different interferences levels brought by them.

Specifically, given a matched BS-subchannel unit $\left( {m,k} \right)$, it prefers to be cohabitated with a new pair $\left( {j',\left( {m',k} \right)} \right)$ with high-quality user $j'$ -- BS $m'$ link and poor-quality user $j'$ -- BS $m$ interference link over subchannel $k$. To depict such mutual effect, we construct a preference matrix ${\textbf{\emph{Z}}}_{m,k}$ of size $(M+1) \times J$ for each BS-subchannel unit $\left( {m,k}\right)$ evaluating all possible matching pairs over subchannel $k$. Each element in ${\textbf{\emph{Z}}}_{m,k}$ is defined as
\begin{equation} \label{preference}
\zeta _{m',j',k}^{\left( {m,k} \right)} = \left. {\left| {h_{m',j',k}^B} \right|^{{\rho _1}{{\lambda _{m'}}} }} \middle/  {\left| {h_{m,j',k}^B} \right|^{{\rho _2}}}\right.,
\end{equation}
where $\rho_1$ and $\rho_2$ are preference parameters. We then say that a BS-subchannel unit $\left( {m,k} \right)$ prefers  $\left( {j_1,\left( {m_1,k} \right)} \right)$ to  $\left( {j_2,\left( {m_2,k} \right)} \right)$ if ${\zeta _{{m_1},{j_1},k}^{\left( m,k\right) }} > {\zeta _{{m_2},{j_2},k}^{\left( m,k\right) }}$, i.e.,
\begin{equation} \label{pairwise_preference}
\left( {{j_1},\left( {{m_1},k} \right)} \right){ \succ _{\left(  {m,k} \right)}}\left( {{j_2},\left( {{m_2},k} \right)} \right) \Leftrightarrow \zeta _{{m_1},{j_1},k}^{\left( m,k\right) } > {\zeta _{{m_2},{j_2},k}^{\left( m,k\right) }},\forall {j_1}{j_2} \in {\cal{J}}_{un},\forall {m_1},{m_2} \ne m,
\end{equation}
where ${\cal{J}}_{un}$ is the set of unmatched users. Each BS-subchannel unit's attitude towards the potential matching pairs is affected by the preference parameters $\rho_1$ and $\rho_2$ in $\left( \ref{preference} \right)$. When $\rho_1 = 0$, the unit has a \emph{pessimistic attitude}, i.e., it only cares to minimize the interference brought by a potential matching pair. When $\rho_2 = 0$, the unit has an \emph{optimistic attitude}, i.e., it only cares to maximize the benefit brought by the new matching pair. Accordingly, $\rho_1 = \rho_2$ reflects a \emph{neutral attitude}.

The traditional preference of each subchannel over different matchings is also introduced. Define the utility of each subchannel $k$ as the sum rate of all BSs over this subchannel, i.e., $R_k^B = \sum\limits_{m = 0}^M {\left( {1 - {\lambda _m}} \right)\sum\limits_{j = 1}^J {{R_{m,j,k}^B}} }$. The preference of subchannel $k$ is then given by
\vspace{-0.2cm}
\begin{equation}
{\Psi _1}{ \succ _k}{\Psi _2} \Leftrightarrow R_k^B\left( {{\Psi _1}} \right) > R_k^B\left( {{\Psi _2}} \right),
\end{equation}
\vspace{-0.1cm}
where $R_k^B\left( {{\Psi _1}} \right)$ is the utility obtained by subchannel $k$ under $\Psi_1$.


Due to the logarithmic-form utility function, the externalities in our formulated problem do not share the additive characteristic with other well-defined one-to-one matchings with externalities~\cite{mumcu2010stable}.


\vspace{-0.3cm}
\subsection{Algorithm Design}

\vspace{-0.1cm}
\subsubsection{Initialization}
We adopt a greedy algorithm to initialize the matching where each subchannel is matched with the most preferred combination of a BS and a user that can accept it. Specifically, each unmatched subchannel $k$ proposes to an unmatched user $j$ and BS $m$ to match with satisfying
\vspace{-0.2cm}
\begin{equation} \label{initialization}
{j^*},{m^*} = \arg \mathop {\max }\limits_{j \in {{\cal{J}}_{un}},0 \le m \le M} {\left| {h_{m,j,k}^B} \right|^2}.
\end{equation}
\vspace{-0.2cm}
If user $j$ is proposed by more than one subchannel, then it selects a unit $\left( {m,k} \right)$ with the largest channel gain from the candidates and rejects the others.

\subsubsection{Propose-and-Reject Operation}~\label{propose_reject}
Based on the defined preference relationship, we introduce the key operation of each matching pair, consisting of one proposing phase and two rejecting phases. Different from the traditional matchings, each matched BS-subchannel unit is allowed to propose to potential matching pairs instead of users.
\begin{itemize}
\item \textbf{BS-subchannel unit proposing:} Each matched BS-subchannel unit $\left( {m,k} \right)$ in the current matching $\Psi$ selects its most preferred matching pair $\left( {j',\left( {m',k} \right)}\right) $ satisfying
    \vspace{-0.2cm}
\begin{equation} \label{pairwise_propose}
j',\left( {m',k} \right) = \arg \mathop {\max }\limits_{j' \in {{\cal{J}}_{un}},m' \in {{\cal{M}}_{un,k}}} \left\{ {\zeta _{m',j',k}^{\left( {m,k} \right)}|{a_{m',j'}} = 1} \right\},
\end{equation}
\vspace{-0.2cm}
where ${\cal{M}}_{un,k}$ is the set of unmatched BSs over subchannel $k$. A set of candidate pairs ${\cal{S}}_k$ consisting of $\left( {j',\left( {m',k} \right)}\right) $ is then constructed for each subchannel $k$.
\item \textbf{Subchannel Rejecting:} Each subchannel $k$ selects one candidate matching pair $\left( j'',\left( {m'',k} \right) \right) $ from ${\cal{S}}_k$ such that
    \vspace{-0.2cm}
\begin{equation} \label{subchannel_reject}
\begin{split}
\left\lbrace  {j'',\left( {m'',k} \right)} \right\rbrace  \cup \Psi &{ \succ _k}\Psi,\\
\left\lbrace  {j'',\left( {m'',k} \right)} \right\rbrace  \cup \Psi &{ \succeq _k}\left\lbrace  {j',\left( {m',k} \right)} \right\rbrace  \cup \Psi ,\forall \left( {j',\left( {m',k} \right)} \right) \in {{\cal{S}}_k},
\end{split}
\end{equation}
\vspace{-0.2cm}
i.e., the matching pair that brings the highest positive utility to subchannel $k$. Other matching pairs in ${\cal{S}}_k$ are then rejected.
\item \textbf{User Rejecting:} If an unmatched user $j''$ is proposed more than once, it only accepts the BS-subchannel unit with the highest utility and rejects all the other proposals. Once a user is matched, it is removed from ${\cal{J}}_{un}$.
\end{itemize}

\subsubsection{Algorithm Description}
The whole matching algorithm for terrestrial user association and subchannel allocation (TUASA) is presented in detail in Table~\ref{Alg_1}. In the initialization step (line 2-12), each subchannel is matched with a combination of one user and one BS with the best channel condition that it can achieve. The following matching process (line 14-33) consists of multiple iterations in each of which the propose-and-reject operation (line 16-29) is performed. The iterations will not stop until no matched BS-subchannel unit would like to propose to the users any more.

\begin{algorithm}[!t]
	\caption{Terrestrial User Association and Subchannel Allocation (TUASA) Algorithm}\label{Alg_1}
	
	\hspace*{-0.02in} {\bf Input:} 
	Sets of users, BSs, and subchannels $\mathcal{J}$, $\mathcal{M}$, and $\mathcal{K}$; coverage matrix $\emph{\textbf{A}}$.\\
	\hspace*{-0.02in} {\bf Output:} 
	A final matching $\Psi^*$.
	\begin{algorithmic}[1]
		\vspace{-0.2cm}
		\State \textbf{Initialization}
		\vspace{-0.2cm}
		\State Record current matching as $\Psi$. Construct $\mathcal{J}_{un} = \cal{J}$ and $\mathcal{M}_{un,k} = \cal{M}$, $\forall 1 \le k \le K$.
		\vspace{-0.2cm}
		\While{at least one subchannel is unmatched}
		\vspace{-0.2cm}
		\State Each unmatched subchannel $k$ proposes to user $j^*$ and BS $m^*$ according to $\left( \ref{initialization} \right)$
		\vspace{-0.2cm}
		\For{each user receiving proposals}
		\vspace{-0.2cm}
		\If{user $j$ is proposed by more than one subchannel}
		\vspace{-0.2cm}
		\State user $j$ selects $\left( m,k \right) $ with the highest channel gain ${\left| {h_{m,j,k}^B} \right|^2}$ from the candidates
		\vspace{-0.2cm}
		\State User $j$ rejects the other proposals
		\vspace{-0.2cm}
		\Else
		\vspace{-0.2cm}
		\State User $j$ is matched with the proposing BS-subchannel unit
		\vspace{-0.2cm}
		\EndIf
		\vspace{-0.2cm}
		\State User $j$ is removed from ${\cal{J}}_{un}$
		\vspace{-0.2cm}
		\EndFor
		\vspace{-0.1cm}
		\EndWhile		
		\State \textbf{Matching Process}
		\vspace{-0.2cm}
		\While{${\cal{J}}_{un} \ne \varnothing$ or some matched BS-subchannel unit still tries to propose}
		\vspace{-0.2cm}
		\State Set ${\cal{S}}_k = \varnothing$ for all $1 \le k \le K$.
		\vspace{-0.2cm}
		\For{each subchannel $k$}
		\vspace{-0.2cm}
		\For{each matched BS-subchannel unit $\left( {m,k} \right)$ in $\Psi$}
		\vspace{-0.2cm}
		\State Propose to its most preferred pair ${\left( {j',\left( {m',k} \right)} \right)}$ according to $\left( \ref{pairwise_propose} \right)$
		\vspace{-0.2cm}
		\State Add ${\left( {j',\left( {m',k} \right)} \right)}$ to ${\cal{S}}_k$.
		\vspace{-0.2cm}
		\EndFor
		\vspace{-0.2cm}
		\State Select one candidate pair ${\left( {j'',\left( {m'',k} \right)} \right)}$ from ${\cal{S}}_k$ according to $\left( \ref{subchannel_reject}\right)$.
		\vspace{-0.2cm}
		\State Remove other candidate pairs from ${\cal{S}}_k$.
		\vspace{-0.2cm}
		\EndFor
		\vspace{-0.2cm}
		\For{each user $j$ included in ${\cal{S}}_k$ ($1 \le k \le K$)}
		\vspace{-0.2cm}
		\If{it is proposed more than once}
		\vspace{-0.2cm}
		\State User $j$ accepts the best  BS-subchannel unit and rejects others.
		\vspace{-0.2cm}
		\Else
		\vspace{-0.2cm}
		\State User $j$ accepts the received proposal.
		\vspace{-0.2cm}
		\EndIf
		\vspace{-0.2cm}
		\State User $j$ is removed from ${\cal{J}}_{un}$.
		\vspace{-0.2cm}
		\State The newly matched BS is removed from the corresponding $\mathcal{M}_{un,k}$.
		\vspace{-0.2cm}
		\EndFor
		\vspace{-0.2cm}
		\EndWhile
		\State \Return the final matching $\Psi^*$
	\end{algorithmic}
\end{algorithm}
\vspace{-0.6cm}
\subsection{Algorithm Analysis}
\vspace{-0.2cm}
\subsubsection{Remark on the number of accessed users}
The propose-and-reject operation offers the optimal strategy to maximize the number of accessed users in the TTO problem since each user $j$ is compulsively served by a BS as long as there exist an unmatched BS-subchannel unit $\left(m,k\right)$ satisfying $a_{m,j} = 1$. In other words, the proposed TUASA algorithm guarantees the maximum number of accessed users while improving the sum rate.
\subsubsection{Stability and Convergence}
As proved in Appendix A, our proposed TUASA algorithm is guaranteed to converge to a final matching after a limited number of iterations.


Due to the complicated interaction relationship between different pairs brought by the co-channel interference, the preference in our formulated problem does not satisfy the substitutability condition, which is usually a sufficient condition for the existence of a stable matching. Therefore, the traditional solution concepts of blocking pair and stability cannot be applied in this case. We then introduce the concept of a blocking pair, which is stricter than the traditional version.

\textbf{Definition 1:} Given a matching $\Psi$, where $\Psi \left( j_1 \right) = \left( {{m_1},{k_1}} \right)$ and $\Psi \left( {m_2,k_2} \right) = {j_2}$ ($k_1$ and $k_2$ are allowed to be equal), we denote the matching where user $j_1$ is matched with $\left( m_2,k_2\right)$ while other pairs are unchanged as $\Psi ' = \Psi \backslash \left\{ {\left( {j_1,\left( {{m_1},{k_1}} \right)} \right),\left( {{j_2},\left( {m_2,k_2} \right)} \right)} \right\} \cup \left\{ {\left( {j_1,\left( {m_2,k_2} \right)} \right),\left( {{j_2},\left( {{m_1},{k_1}} \right)} \right)} \right\}$. The pairs $\left( {j_2,\left( {m_1,k_1} \right)} \right)$ and $\left( {j_1,\left( {m_2,k_2} \right)} \right)$ are \emph{individual-rational blocking pairs} if i) $R_{m',j',k_i}^B\left( {\Psi '} \right) > R_{m',j',k_i}^B\left( \Psi  \right)$,\hspace{0.2cm}$\forall j',\left( {m',k_i} \right) \in \Psi$, $i = 1,2$; ii) $R_{{m_1},{j_2},{k_1}}^B\left( {\Psi '} \right) > R_{m_1,j_1,{k_1}}^B\left( \Psi  \right)$ and $R_{{m_2},{j_1},{k_2}}^B\left( {\Psi '} \right) > R_{m_2,j_2,{k_2}}^B\left( \Psi  \right)$. In other words, the blocking pairs $\left( {j_1,\left( {m_2,k_2} \right)} \right)$ and $\left( {j_2,\left( {m_1,k_1} \right)} \right)$ can bring higher utility to all existing pairs with respect to subchannels $k_1$ and $k_2$ in $\Psi$.

Given the above definition, we discuss the stability and equilibrium based on different attitudes of the BS-subchannel units. The following statements are proved in Appendix B.
\begin{itemize}
	\item When $\rho_2 = 0$, we have the \emph{group stability} concept in the sense that $j_1$, $j_2$, $\left( m_1,k_1\right) $, and $\left( m_2, k_2\right) $ are considered as a group. In other words, there does not exist individual-rational blocking pairs satisfying condition ii) in Definition 1.
	\item When $\rho_2 \ne 0$, instead of the group stability, we have the concept of equilibrium. When there is no individual-rational blocking pair in a matching, any new matching pair $\left( {j,\left( {m,k} \right)} \right)$ cannot improve the utility of the $\left( {m,k} \right)$ unit without compromising the utility of other matched BS-subchannel units. The final matching then reaches an \emph{equilibrium point}.
\end{itemize}

\subsubsection{Computational Complexity} \label{complexity_alg1}
The maximum number of iterations in the initialization phase is $J$, and the number of outer iterations in the matching phase is also proportional to $J$. For the worst case, in each outer iteration all matched pairs propose to the same user such that only one new matching pair is formed. The total number of outer iterations in this case is $J$. For the best case, in each iteration a new matching pair is accepted over a subchannel, i.e., $K$ new pairs are recorded. Since there are $J$ users in total, the number of required iterations is $\left\lceil {J/K} \right\rceil$. In practice, the iteration number varies between $\left\lceil {J/K} \right\rceil$ and $J$.


%

\vspace{-0.25cm}
\section{Algorithm Design for LEO-based Backhaul Capacity Maximization}
In this section, we convert the LBCO problem in $\left(\ref{problem2} \right) $ into a many-to-one matching problem with externalities. To depict the multi-connectivity and angle-sensitive nature of the UD-LEO networks, a novel swap matching algorithm with continuous power control (SMPC) is proposed.
\vspace{-0.75cm}
\subsection{Matching Problem Formulation}
\vspace{-0.1cm}
\subsubsection{Definitions}
Consider the set of TSTs, satellites, and available subchannels as ${\cal{M}}'$, $\cal{N}$, and $\cal{Q}$. We assume that each satellite $n$ and subchannel $q$ form a satellite-subchannel (SS) unit $\left( n,q\right)$. A matching $\Phi$ is defined as a mapping between ${\cal{M}}'$ and $\cal{N} \times \cal{Q}$ such that for each TST $m$ and each SS unit $\left( n,q\right)$, we have: i) $\Phi \left( {n,q} \right) = m$ if and only if $\left( {n,q} \right) \in \Phi \left( m \right)$; ii) $\Phi \left( m \right) \subseteq (\cal{N} \times \cal{Q}) \cup \varnothing$ and $\left| {\Phi \left( m \right)} \right| \le {N_r}$; iii) $\Phi \left( {n,q} \right) \in {\cal{M}}'$ and $\left| {\Phi \left( {n,q} \right)} \right| \le 1$. Specifically, each matching pair is denoted by ${\left( {m,\left( {n,q} \right)} \right)_{p_{m,n}^T}}$ , where $p_{m,n}^T$ is the transmit power of TST $m$ over this link, satisfying constraint (9g). We aim to find a matching such that the weighted capacity $\sum\nolimits_{m = M' + 1}^M {{\lambda _m}} {{\tilde C}_m}$ can be maximized. The equivalent capacity can then be obtained by solving equations in $\left( \ref{equation_set}\right)$.
\subsubsection{Preference Relation} \label{preference_LEO}


Following Section~\ref{TTO_preference}, the influence brought by a potential pair $\left( m', \left( n',q\right) \right)$ with respect to the existing matched SS unit $\left( n,q\right)$ is re-defined as
\begin{equation} \label{TST_preference_matrix}
w_{m',n',q}^{\left( {n,q} \right)} = \left.{v_{m',n'}}{{{\left[ {G_{m',n'}^{m',n'}{{\left| {h_{m',n',q}^T} \right|}^2}} \right]}^{{\rho _1}}}}\middle/{{{{\left[ {G_{m',n}^{m',n'}{{\left| {h_{m',n,q}^T} \right|}^2}} \right]}^{{\rho _2}}}}}\right..
\end{equation}
Therefore, the preference relation $\succ_{\left( n,q\right)}$ can be given by
\vspace{-0.2cm}
\begin{equation} \label{TST_preference}
\left( {{m_1},\left( {{n_1},q} \right)} \right){ \succ _{\left( {n,q} \right)}}\left( {{m_2},\left( {{n_2},q} \right)} \right) \Leftrightarrow w_{{m_1},{n_1},q}^{\left( {n,q} \right)} > w_{{m_2},{n_2},q}^{\left( {n,q} \right)},\forall {m_1},{m_2},\forall {n_1},{n_2} \ne n.
\end{equation}
\vspace{-0.2cm}
Each subchannel's preference over different matchings is similar to that in Section~\ref{TTO_preference}. The utility of subchannel $q$ is defined as $R_q^T = \sum\nolimits_{m = M' + 1}^M {{\lambda _m}} \sum\nolimits_{n = 1}^N {R_{m,n,q}^T}$ in this case.
\subsubsection{Remark on the New Problem}
Two new challenges have been posed compared to the TTO problem, rendering most existing matching algorithms~\cite{manlove2013algorithmics,di2016sub,di2017non} as well as Algorithm~\ref{Alg_1} not suitable any more, as illustrated below.

First, there lacks a mechanism for transmit power adjustment in each project-and-reject operation as shown in Section~\ref{propose_reject}. In addition, the complete preference list based on $\left( \ref{TST_preference_matrix}\right)$ over all possible power levels is very difficult to construct.

Second, due to the angular sensitivity, the group stability in Algorithm~\ref{Alg_1} cannot be guaranteed any more. Typically, in the matching $\Psi$ obtained from Algorithm~\ref{Alg_1}, the interference caused by user $j$ to a given BS $m$ is irrelevant of its matched BS. However, in the LBCO problem, as long as TST $m$ switches its matched satellites from $n_1$ to $n_2$, the co-channel interference brought to any other satellite $n'$ varies due to different off-axis antenna gains, i.e., $G_{m,n'}^{m,{n_2}} \neq G_{m,n'}^{m,{n_1}} \neq G_{m,n_2}^{m,{n_2}}$. By comparing equations $\left( \ref{signal_model} \right)$ and $\left( \ref{signal_LEO} \right)$, we can infer that the group stability does not hold in the LBCO problem even if $\rho_2 = 0$, implying that the propose-and-reject operation fails to depict such dynamic matching structure change over each subchannel in the LBCO problem.

\vspace{-0.5cm}
\subsection{Algorithm Design}
\vspace{-0.2cm}
We introduce the swap matching performed by each TST as below. Generally speaking, in a swap operation, a TST tends to swap its matches with another TST while keeping other TSTs' strategies unchanged.

\textbf{Definition 2:} Given a matching $\Phi$  with two existing matching pairs ${\left( {{m_1},\left( {{n_1},{q_1}} \right)} \right)_{p_{{m_1},{n_1},q_1}^T}}$ and ${\left( {{m_2},\left( {{n_2},{q_2}} \right)} \right)_{p_{{m_2},{n_2},q_2}^T}}$, a swap matching is defined as
\begin{equation} \label{swap_matching}
\begin{split}
\Phi _{{m_1},\left( {{n_1},{q_1}} \right)}^{{m_2},\left( {{n_2},{q_2}} \right)} = \Phi \backslash \!\left\{ {{{\left( {{m_1},\left( {{n_1},{q_1}} \right)} \right)}_{p_{{m_1},{n_1},q_1}^T}}\!,{{\left( {{m_2},\left( {{n_2},{q_2}} \right)} \right)}_{p_{{m_2},{n_2},q_2}^T}}} \right\} \cup \\
\left\{ {{{\left( {{m_2},\left( {{n_1},{q_1}} \right)} \right)}_{p_{{m_2},{n_1},q_1}^T}}\!,{{\left( {{m_1},\left( {{n_2},{q_2}} \right)} \right)}_{p_{{m_1},{n_2},q_2}^T}}} \right\},
\end{split}
\end{equation}
Both the TSTs and SS units are allowed to be virtual. There is no physical meaning for a virtual TST or SS unit and the corresponding utility is zero.

\textbf{a) Power Allocation in Swap Matchings:} Based on this generalized definition, we discuss the power control strategy for five typical types of swap matchings. Other cases not mentioned here can be classified into one of the following cases.

\emph{Type-1 (TST $m_2$ and SS unit $\left( n_1, q_1 \right)$ are virtual):} The swap matching degrades to TST $m_1$ proposing to the unmatched SS unit ${\left( {{n_2},{q_2}} \right)}$ and ${p_{{m_2},{n_2},q_2}^T} = 0$. Note that by allocating power to the TST $m_1$ -- satellite $n_2$ link, the SS units in $\Phi \left( {{m_1}} \right)$ may be influenced since the allocated power needs to be adjusted to satisfy the power constraint. Therefore, different from the traditional swap matching, more than two matching pairs may be involved during the swap matching in the LBCO problem.

To evaluate how the power control influences other SS units in $\Phi \left( {{m_1}} \right)$, we consider a pair $\left( {{m_1},\left( {{n},{q}} \right)} \right)$ with $\left( n, q\right) \in \Phi \left( {{m_1}} \right)$. The utility of subchannel $q$ is rewritten as a function of ${p_{{m_1},{n},q}^T}$,
\begin{equation}
R_{{q}}^T\left( {p_{{m_1},{n},q}^T} \right) = {\lambda _{{m_1}}}{\log _2}\left( {1 + p_{{m_1},{n},q}^T \cdot \frac{{G_{{m_1}{n}}^{{m_1},{n}}{{\left| {h_{{m_1},{n},{q}}^T} \right|}^2}}}{{{I_{{n},{q}}} + {\sigma ^2}}}} \right) \nonumber
\end{equation}
\begin{equation}
\quad \quad + \sum\limits_{m'} {\lambda _{{m'}}} {\sum\limits_{n' \ne {n_i}} {{{\log }_2}\left( {1 + \frac{{{S_{m',n',{q}}}}}{{p_{{m_1},{n},q}^TG_{{m_1},n'}^{{m_1},{n}}{{\left| {h_{{m_1},n',{q}}^T} \right|}^2} + {I_{n',{q}}} + {\sigma ^2}}}} \right)} },
\end{equation}
where ${{I_{{n},{q}}}}$ (or ${{I_{n',{q}}}}$) is the interference received by satellite $n_i$ (or $n'$) over subchannel $q_t$ and ${{S_{m',n',{q}}}}$ is the signal strength of the TST $m'$ -- satellite $n'$ link. All these three items are irrelevant of ${p_{{m_1},{n},q}^T}$.

Define the negative gradient of $R_{{q}}^T\left( {p_{{m_1},{n},q}^T} \right)$ as $f\left( {p_{{m_1},{n},q}^T} \right) =  - {\nabla _{p_{{m_1},{n},q}^T}}\left( {R_{{q}}^T} \right)$. It is observed that the larger $f\left( {p_{{m_1},{n},q}^T} \right)$ is, the smaller influence $\left( {{m_1},\left( {{n_i},{q_t}} \right)} \right)$ suffers from the reduction of its transmit power. Therefore, we select a pair $\left( {{m_1},\left( {{n^*},{q^*}} \right)} \right)$ from $\Gamma \left( {{m_1}} \right)$ the least affected by the swap matching, i.e.,
\vspace{-0.2cm}
\begin{equation} \label{gradient_power}
\left( {{n^*},{q^*}} \right) = \arg \mathop {\max }\limits_{\left( {n,q} \right) \in \Gamma \left( {{m_1}} \right)} f\left( {p_{{m_1},n,q}^T} \right).
\end{equation}
\vspace{-0.1cm}
To perform the swap matching, we consider maximizing the total utility of the involved subcahnnels via the power control of TST $m_1$. For convenience, we denote the available power budget of this swap matching as $P_{{m_1}}^{budget} = {p_{{m_1}}}\left( \Phi  \right) + p_{{m_1},{n^*},{q^*}}^T\left( \Phi  \right)$, where ${p_{{m_1}}}\left( \Phi  \right)$ is the unallocated power of TST $m_1$. The power control problem for $m_1$ can be formulated as
\begin{equation} \label{power_problem1}
\begin{split}
(\underline{PC1}:) & \mathop {\max }\limits_{\left\{ {p_{{m_1},{n_2},{q_2}}^T,p_{{m_1},{n^*},{q^*}}^T} \right\}} R_{{q_2}}^T\left( {p_{{m_1},{n_2},q_2}^T} \right) + R_{{q^*}}^T\left( {p_{{m_1},{n^*},{q^*}}^T} \right)\\
\emph{s.t.}\ & p_{{m_1},{n_2},{q_2}}^T + p_{{m_1},{n^*},{q^*}}^T \le P_{{m_1}}^{budget}.
\end{split}
\end{equation}
Specifically, when $q^* = q_2$, the objective function $R_{{q^*}}^T\left( {p_{{m_1},{n_2},{q^*}}^T,p_{{m_1},{n^*},{q^*}}^T} \right)$ becomes
\begin{equation} \label{PC2_objective}
{\lambda _{{m_1}}}\left[ {{{\log }_2}\left( {1 + \frac{{p_{{m_1},{n^*},{q^*}}^TA}}{{{I_{{n^*},{q^*}}} + p_{{m_1},{n_2},{q^*}}^TB + {\sigma ^2}}}} \right) + {{\log }_2}\left( {1 + \frac{{p_{{m_1},{n_2},{q^*}}^TC}}{{{I_{{n_2},{q^*}}} + p_{{m_1},{n^*},{q^*}}^TD + {\sigma ^2}}}} \right)} \right] \nonumber
\end{equation}
\vspace{-0.2cm}
\begin{equation}
+  \sum\limits_{m' \ne {m_1}} {{\lambda _{m'}}\sum\limits_{n' \ne {n^*} \ne {n_2}} {{{\log }_2}\left( {1 + \frac{{{S_{m',n',{q^*}}}}}{{{I_{n',{q^*}}} + p_{{m_1},{n^*},{q^*}}^T{F_{m',n'}} + p_{{m_1},{n_2},{q^*}}^T{G_{m',n'}} + {\sigma ^2}}}} \right)} },
\end{equation}
where $A \sim D$, $F_{m',n'}$, and $G_{m',n'}$ are all channel-relevant constants, and the detailed forms are omitted here. The power control problem of TST $m_1$ is then formulated as:
\vspace{-0.2cm}
\begin{equation} \label{power_problem2}
\begin{split}
(\underline{PC2}:) & \mathop {\max }\limits_{\left\{ {p_{{m_1},{n_2},{q^*}}^T,p_{{m_1},{n^*},{q^*}}^T} \right\}} R_{{q^*}}^T\left( {p_{{m_1},{n_2},{q^*}}^T,p_{{m_1},{n^*},{q^*}}^T} \right)\\
\emph{s.t.}\ & p_{{m_1},{n_2},{q^*}}^T + p_{{m_1},{n^*},{q^*}}^T \le P_{{m_1}}^{budget}.
\end{split}
\end{equation}
The above two power control problems can be easily solved, as presented in Appendix C.

\emph{Type-2 ($\left( {{m_2},\left( {{n_2},{q_2}} \right)} \right)$ is a virtual pair):} The swap matching degrades to the case where TST $m_1$ considers to withdraw some power resources allocated to the TST $m_1$ -- satellite $n_1$ link over subchannel $q_1$. The power control problem for TST $m_1$ is formulated as
\begin{equation} \label{power_problem3}
\begin{split}
(\underline{PC3}:) &\mathop {\max }\limits_{{p_{{m_1},{n_1},{q_1}}}} R_{{q_1}}^T\left( {p_{{m_1},{n_1}}^T} \right) \\
\emph{s.t.}\ & 0 \le {p_{{m_1},{n_1},q_1}} \le {p_{{m_1},{n_1},q_1}}\left( \Phi  \right).
\end{split}
\end{equation}
The solution can be found by traversing the extreme points and boundary points of the objective function, as illustrated in Appendix C.

\emph{Type-3 ($m_1 = m_2$):} In this case, TST $m_1$ adjusts its power allocated to two links. The power budget of $m_1$ is $P_{{m_1}}^{budget} = p_{{m_1},{n_1},q_1}^T\left( \Phi  \right) + p_{{m_1},{n_2},q_2}^T\left( \Phi  \right) + {p_{{m_1}}}\left( \Phi  \right)$. When $q_1 \ne q_2$, the power control problem follows a PC1 fashion shown in $\left( \ref{power_problem1} \right)$ with the objective function $R_{{q_2}}^T\left( {p_{{m_1},{n_2},q_2}^T} \right) + R_{{q_1}}^T\left( {p_{{m_1},{n_1},q_1}^T} \right)$. In contrast, when $q_1 = q_2 = q$, this becomes a PC2-type problem as shown in $\left( \ref{power_problem2} \right)$ with the objective function $R_q^T\left( {p_{{m_1},{n_2},q}^T,p_{{m_1},{n_1},q}^T} \right)$.

\emph{Type-4 (No nodes are virtual or identical):} TSTs $m_1$ and $m_2$ swap their matches over different subchannels with their power budgets unchanged. We evaluate the total utility of two involved subchannels such that each TST separately solves a PC3-type problem. The power budget for TST $m_1$ is $P_{{m_1}}^{budget} = p_{{m_1},{n_1},q_1}^T\left( \Phi  \right) + {p_{{m_1}}}\left( \Phi  \right)$ and the objective function is $R_{{q_2}}^T\left( {p_{{m_1},{n_2},q_2}^T} \right)$ where the TST $m_2$ -- satellite $n_2$ link has been removed. A symmetric problem can be formulated for TST $m_2$. After solving the above two PC3 problems, the total utility is obtained by $R_{{q_2}}^T\left( {p_{{m_1},{n_2},q_2}^{T,*}} \right) + R_{{q_1}}^T\left( {p_{{m_2},{n_1},q_1}^{T,*}} \right)$, where ${p_{{m_1},{n_2},q_2}^{T,*}}$ and ${p_{{m_2},{n_1},q_1}^{T,*}}$ are the solutions for two PC3 problems, respectively.

\emph{Type-5 (No nodes are virtual and only $q_1 = q_2 = q$):} TSTs $m_1$ and $m_2$ swap their matches over subchannel $q$ with their power budgets unchanged. This directly degrades to a PC2 problem where the constraints are separated as those in the Type-4 swap matching.

\textbf{b) Feasibility and Validity of A Swap Matching:} To maximize the total utility of the matching, each potential swap matching needs to be evaluated before being executed.

\textbf{Definition 3:} A swap matching ${\Phi _{{m_1},\left( {{n_1},{q_1}} \right)}^{{m_2},\left( {{n_2},{q_2}} \right)}}$ is \emph{feasible} if and only if all constraints in $\left( \ref{problem2}\right)$ are not violated, i.e., i) $\left| {{\theta _{{m_1},{n_2}}} - {\theta _{{m_1},{n_i}}}} \right| \ge {\theta _{th}},\forall {n_i} \in \Phi \left( {{m_1}} \right),{n_i} \ne {n_2}$ and $\left| {{\theta _{{m_2},{n_1}}} - {\theta _{{m_2},{n_{i'}}}}} \right| \ge {\theta _{th}},\forall {n_{i'}} \in \Phi \left( {{m_2}} \right),{n_{i'}} \ne {n_1}$; ii) $\left| {\Phi \left( {{m_1}} \right)} \right| \le {N_r}$ and $\left| {\Phi \left( {{m_2}} \right)} \right| \le {N_r}$; iii) $\left| {\Phi \left( {n_1,q_1} \right)} \right| \le 1$ and $\left| {\Phi \left( {n_2,q_2} \right)} \right| \le 1$ still stand after the swap matching. Based on the power strategy of two TSTs, a swap matching can only be \emph{approved} if the total utility of the corresponding subchannels is improved, i.e.,
\begin{equation} \label{swap_approve}
R_{{q_1}}^T\left( {\Phi _{{m_1},\left( {{n_1},{q_1}} \right)}^{{m_2},\left( {{n_2},{q_2}} \right)}} \right) + R_{{q_2}}^T\left( {\Phi _{{m_1},\left( {{n_1},{q_1}} \right)}^{{m_2},\left( {{n_2},{q_2}} \right)}} \right) > R_{{q_1}}^T\left( \Phi  \right) + R_{{q_2}}^T\left( \Phi  \right).
\end{equation}
For case 1, $\left({{m_2},\left( {{n_2},{q_2}} \right)} \right)$ is replaced by $\left({{m_1},\left( {{n^*},{q^*}} \right)} \right)$.

\textbf{c) Pruning Procedure for Swap Matching Candidate Selection: } In traditional swap matching algorithms, every two matching pairs are searched for finding an approved swap matching as long as the matching structure changes. Each time it involves optimizing and comparing the utility before and after the candidate swap matching, thereby leading to a high complexity.

To address this issue, we add a gradient-based pruning procedure for preprocessing before the swap matching. Given a matching $\Phi$, we construct a gradient matrix $\emph{\textbf{W}}_m$ of size $N \times Q$ for each TST $m$ where the element $w^m_{n,q}$ is defined as
\begin{equation} \label{gradient_variable}
w_{n,q}^m =
\begin{cases}
- {\nabla _{p_{m,n}^T}}R_q^T\left( {p_{m,n,q}^T} \right),  &~ \mbox{if $\left( {m,\left( {n,q} \right)} \right) \in \Phi$}, \\
- {\nabla _{p_{m,n}^T}}R_q^T\left( 0^+ \right),  &~ \mbox{otherwise}.
\end{cases}
\end{equation}
The gradient matrix is updated after every approved swap matching ${\Phi _{{m_1},\left( {{n_1},{q_1}} \right)}^{{m_2},\left( {{n_2},{q_2}} \right)}}$. Instead of updating all elements, only those involving $q_1$ and $q_2$ need to be updated. The candidate swap matching can then be selected according the following pruning rules:
\begin{itemize}
	\item Type-1 (or Type-2) swap matching: TST $m_1$ should satisfy $w_{{n_2},{q_2}}^{{m_1}} < 0$ (or $w_{{n_1},{q_1}}^{{m_1}} > 0$);
	\item Type-3 swap matching: TST $m_1$ should satisfy $w_{{n_2},{q_2}}^{{m_1}} \ne w_{{n_1},{q_1}}^{{m_1}}$. Specifically, when $\left(n_2, q_2 \right)$ is an unmatched SS unit, it is required that $w_{{n_2},q_2}^{{m_1}} < 0$ and $w_{{n_2},q_2}^{{m_1}} < w_{{n_1},q_1}^{{m_1}}$;
	\item Type-4 swap matching: it is required that $w_{{n_2},q_2}^{{m_1}} > 0$ and $w_{{n_1},q_1}^{{m_2}} > 0$ cannot hold at the same time ($w_{{n_2},q_2}^{{m_1}}$ and $w_{{n_1},q_1}^{{m_2}}$ need to be updated first based on ${\Phi _{{m_1},\left( {{n_1},{q_1}} \right)}^{{m_2},\left( {{n_2},{q_2}} \right)}}$).
\end{itemize}
By pruning those matching pairs violating the above rules, we do not have to traverse all potential swap matchings. Since updating the gradient matrix is more convenient than solving the power control problems (PC1, or PC2, or PC3), the computation task can be largely reduced.

\textbf{d) Algorithm Description:} Given the above definitions, we propose a swap matching algorithm with power control (SMPC algorithm) for solving the LBCO problem, as shown in detail in Algorithm~\ref{Alg_2}. For initialization, Alg~\ref{Alg_1} is adopted where each TST is only allowed to match one SS unit given fixed power ${P^T}/{N_r}$. The preference relation defined in Section~\ref{preference_LEO} is utilized. The following swap matching process contains multiple iterations in each of which an approved swap matching is found and executed. The iterations will not stop until there is no approved swap matching in the current matching.

\begin{algorithm}[!t]
	\caption{Swap Matching Algorithm with Power Control (SMPC) for solving LBCO problem}\label{Alg_2}
	
	\hspace*{-0.02in} {\bf Input:} 
	Sets of TSTs, satellites, and subchannels ${\mathcal{M}}'$, $\mathcal{N}$, and $\mathcal{Q}$; constraint matrix $\emph{\textbf{V}}$.\\
	\hspace*{-0.02in} {\bf Output:} 
	A swap stable matching $\Phi^*$.
	\begin{algorithmic}[1]
		\vspace{-0.2cm}
		\State \textbf{Initialization}
		\vspace{-0.2cm}
		\State Allocate a fixed transmit power ${P_T}/{N_r}$ to each TST $m$.
		\vspace{-0.2cm}
		\State Perform Algorithm~\ref{Alg_1} to obtain a one-to-one matching $\Phi$.
		\vspace{-0.2cm}
		\State Construct a gradient matrix ${\emph{\textbf{W}}}_m$ for each TST $m$.
		\vspace{-0.2cm}
		\State Denote a virtual TST, satellite, and subchannel as $m_0$, $n_0$, and $q_0$, respectively.
		\vspace{-0.2cm}
		\State Set $p_{{m_1}}^T\left( \Phi  \right) = \left( {1 - 1/{N_r}} \right){P_T}$ for each matched TST $m_1$.
		\vspace{-0.2cm}
		\State Set $p_{{m_2}}^T\left( \Phi  \right) = {P_T}$ for each unmatched TST $m_2$.	
		\vspace{-0.2cm}
		\State \textbf{Swap Matching Process}
		\vspace{-0.2cm}
		\Repeat
		\vspace{-0.2cm}
		\State Select a pair $\left( {m,\left( {n,q} \right)} \right) \notin \Phi$ satisfying pruning rule 1.
		\vspace{-0.15cm}
		\If{$\left| {\Phi \left( m \right)} \right| < {N_r}$ and $\Phi _{{m_0},\left( {n,q} \right)}^{m,\left( {{n_0},{q_0}} \right)}$ is feasible}
		\vspace{-0.15cm}
		\State Execute this swap matching if it is approved according to Definition 3.
		\vspace{-0.15cm}
		\State Set $\Phi _{{m_0},\left( {n,q} \right)}^{m,\left( {{n_0},{q_0}} \right)}$ and update $p_{{m}}^T\left( \Phi  \right)$.
		\vspace{-0.15cm}
		\State Update $w_{n',q}^{m'}$, $\forall n' \in {\cal{N}},m' \in {\cal{M}}'$.
		\vspace{-0.15cm}
		\EndIf
		\vspace{-0.15cm}
		\State Select a pair $\left( {m,\left( {n,q} \right)} \right) \in \Phi$ satisfying pruning rule 2.
		\vspace{-0.15cm}
		\If{$\Phi _{{m_0},\left( {{n_0},{q_0}} \right)}^{m,\left( {n,q} \right)}$ is feasible and can be approved}
		\vspace{-0.15cm}
		\State Execute this swap matching and set $\Phi  = \Phi _{{m_0},\left( {{n_0},{q_0}} \right)}^{m,\left( {n,q} \right)}$.
		\vspace{-0.15cm}
		\State Update $p_{{m1}}^T\left( \Phi  \right)$ and $w_{n',q}^{m'}$, $\forall n' \in {\cal{N}},m' \in {\cal{M}}'$.
		\vspace{-0.15cm}
		\EndIf
		\vspace{-0.2cm}
		\State Select two pairs $\left( {{m_1},\left( {{n_1},{q_1}} \right)} \right),\left( {{m_2},\left( {{n_2},{q_2}} \right)} \right) \in \Phi$.
		\vspace{-0.2cm}
		\If{they fall into Type-3 (or 4, 5) swap matching satisfying the pruning rule}
		\vspace{-0.15cm}
		\If{$\Phi _{{m_1},\left( {{n_1},{q_1}} \right)}^{{m_2},\left( {{n_2},{q_2}} \right)}$ is feasible and can be approved according to Definition 3}
		\vspace{-0.15cm}
		\State Execute this swap matching and set $\Phi  = \Phi _{{m_1},\left( {{n_1},{q_1}} \right)}^{m_2,\left( {n_2,q_2} \right)}$.
		\vspace{-0.15cm}
		\State Update $p_{{m_1}}^T\left( \Phi  \right)$, $p_{{m_2}}^T\left( \Phi  \right)$, $w_{n',q_1}^{m'}$, and $w_{n',q_2}^{m'}$, $\forall n' \in {\cal{N}},m' \in {\cal{M}}'$.
		\vspace{-0.15cm}
		\EndIf
		\vspace{-0.2cm}
		\EndIf
		\Until{the total utility cannot be improved by any swap matching}
			\vspace{-0.2cm}
		\State \Return the final matching $\Phi^*$
	\end{algorithmic}
\end{algorithm}
\vspace{-0.2cm}
\subsection{Stability, Convergence, and Complexity}
\vspace{-0.1cm}
\subsubsection{Convergence}
From Definition 3, each approved swap matching can bring higher utility to the whole system. Note that there exists an upper bound of the weighted capacity given ${\boldsymbol{\lambda}}$ due to the limited resources. Therefore, we can always find an approved matching after which the total utility does not increase any more. The convergence of the proposed SMPC algorithm can then be guaranteed.

\subsubsection{Stability and Equilibrium}
We present the definition of a \emph{swap-stable} matching as below.

\textbf{Definition 4:} A matching $\Phi$ is swap-stable if there is no feasible and approved swap matching that can further improve the total utility.

According to Algorithm~\ref{Alg_2}, the matching process will not stop unitl there is no approved swap matching in the current matching. Therefore, in the final matching, either there is no feasible swap matching, or the remained swap matching cannot improve the utility of the corresponding subchannels. This is naturally a swap-stable matching based on Definition 4. To further analyze the equilibrium of Algorithm~\ref{Alg_2}, we present the following remark, as proved in Appendix D.

\textbf{Remark:} When the TSTs are close enough to each other, i.e., they can be represented by the same geographic coordinates, the final matching is also an equilibrium when $\rho_2 = 0$.
\subsubsection{Complexity} \label{SMPT_complexity}
Since the complexity of the initialization phase is similar to that in Section~\ref{complexity_alg1}, we focus on analyzing the complexity of the swap matching process. Since the complexity has a positive correlation with the number of potential swap matchings in a matching structure $\Phi$, we focus on the latter one to evaluate the complexity. As proved in Appendix E, in the worst case with $\Phi$, potentially we have $NQ\left( {M - M'} \right)$ type-1 (or 2) swap matchings and ${N_r}\left( {{N_r} - 1} \right)\left( {M - M'} \right)/2$ type-3 swap matchings. The number of potential type-4 and type-5 swap matchings is ${N_r}\left( {M - M' - 1} \right)/2 \cdot \max \left\{ {2NQ - \left( {M - M'} \right){N_r},\left( {M - M'} \right){N_r}} \right\}$. In practice, the number of traversed swap matchings after pruning is much smaller than that of the above case, as will shown in Fig~\ref{complexity1}.

For reference, the complexity of the random matching algorithm is $O\left( M \right)$, and that of the greedy algorithm is $O\left( M^2 \right)$. 

\section{Simulation Results}
In this section, we evaluate the performance of our proposed LITS scheme. For comparison, the following schemes are also performed:
\begin{itemize}
	\item \textbf{Ideal backhaul:} We set ideal backhaul for the macro cell and all small cells such that the sum rate obtained in this case is regarded as the upper bound.
	\item \textbf{Traditional terrestrial heterogeneous (TTH) networks:} All small cells are traditionally backhauled via wireless or wired links with limited capacity.
	\item \textbf{Non-integrated terrestrial-satellite (NITS) networks:} This is the sum of the outcomes from two independent networks: 1) the traditional terrestrial network where $J$ users access $M'$ TSCs; 2) the LEO-based network where $J$ users access $M - M'$ LSCs.
\end{itemize}
In our simulation, we set the radius of the macro cell as 1000 m and that of each small cell as 200 m. The transmit power of each user is set as 23 dBm, the data generation rate of each user is 3000 bytes/s, the noise density for terrestrial communications is -174 dBm/Hz, the bandwidth for C-band communications is 20 MHz and that for Ka-band communications is 400 MHz. We set the numbers of subchannels as $K = 10 \sim 20$ and $Q = 10 \sim 15$ in the sense that the satellite networks adopt the broadband communications. The small-scale fading over Ka-band is modeld as Rician fading with the $K$ parameter set as 7, and that over C-band is modeled as the Rayleigh fading. We adopt the Umi path loss model in~\cite{series2009guidelines} for the terrestrial networks and the free-space path loss model for the satellite networks. The backhaul capacities of the macro cell and each TSC are 150Mbps and 20 $\sim$ 30Mbps, respectively. The numbers of TSCs and LSCs are both 25, and the number of LEO satellites is 8. We set the coordinates of satellites according to the satellite tool kit~\cite{satellite_tool} such that the heights of satellites vary from 600km to 1200km. The coordinate of the MBS is set as $\left( {116.38^ \circ} E, {39.92^ \circ}N\right)$. The number of equipped antennas~\cite{OneWeb} for each TST is $N_r = 2$. Other simulation parameters are set according to~\cite{3GPP2017}, as listed in Table~\ref{table_simulation}.
\begin{table}
	\begin{center}
		\caption{Major Simulation Parameters}
		\label{table_simulation}
		\begin{tabular}{|l|l|}
			\hline
			\bf{Parameter} & \bf{Value}\\
			\hline Maximum transmit power of each TST &2W\\
			\hline Antenna type of the TST & 60cm equivalent aperture diameter\\
			\hline Noise figure over Ka band &1.2dB\\
			\hline Ka-band carrier frequency &30GHz\\
			\hline Antenna gain &43.3dBi\\
			\hline Iteration parameter for Lagrangian-based framework  & $10^{-7}$\\
            \hline G/T parameter of each satellite & 18.5 dB/K\\
            \hline Elevation angle~\cite{SpaceX} & 35$^ \circ$ \\
			\hline
		\end{tabular}
	\end{center}
\end{table}

\begin{figure}
\centering
\subfigure[]{
\label{complexity} 
\includegraphics[width=3.2in]{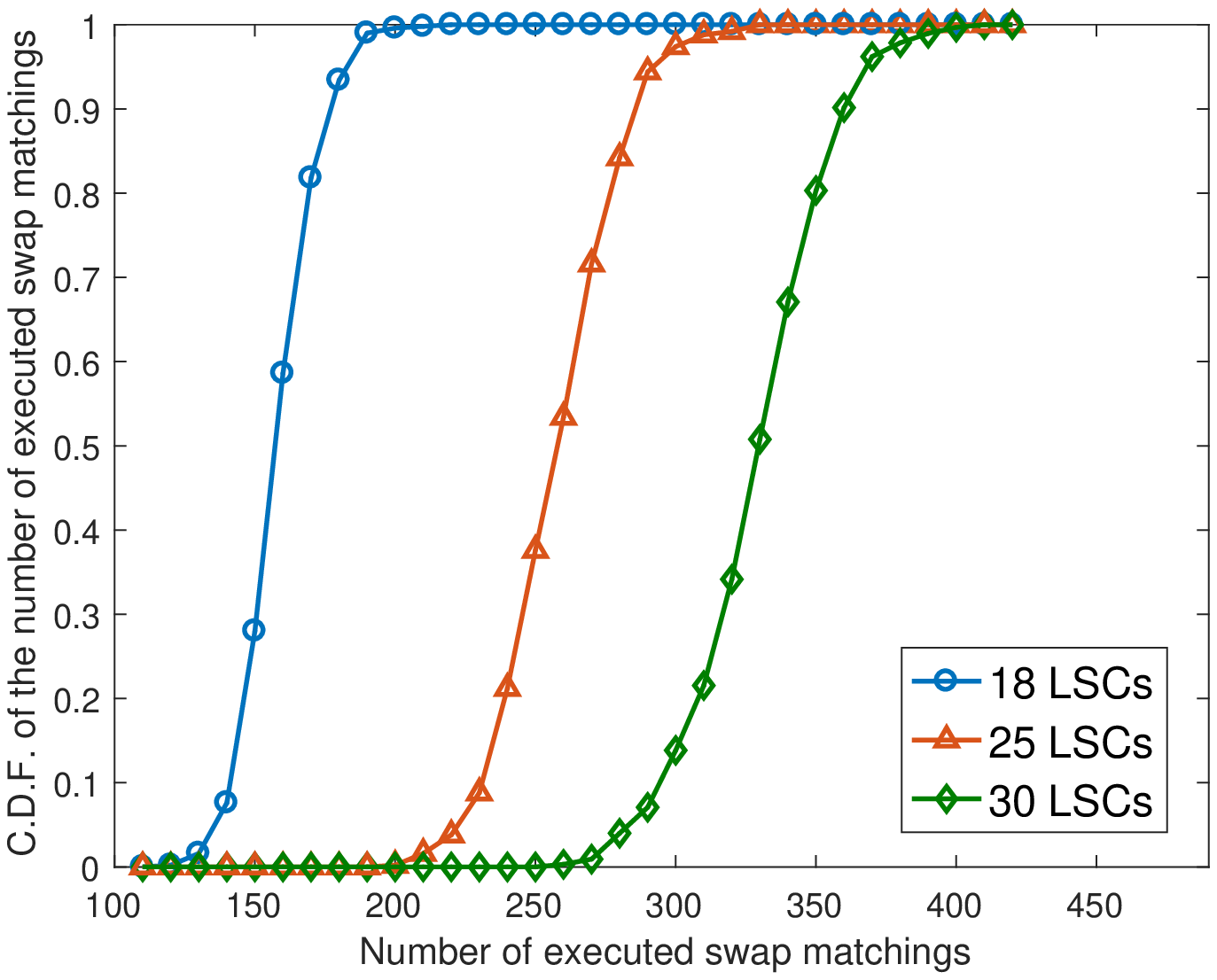}}
\hspace{-0.2in}
\subfigure[]{
\label{complexity1} 
\includegraphics[width=3.2in]{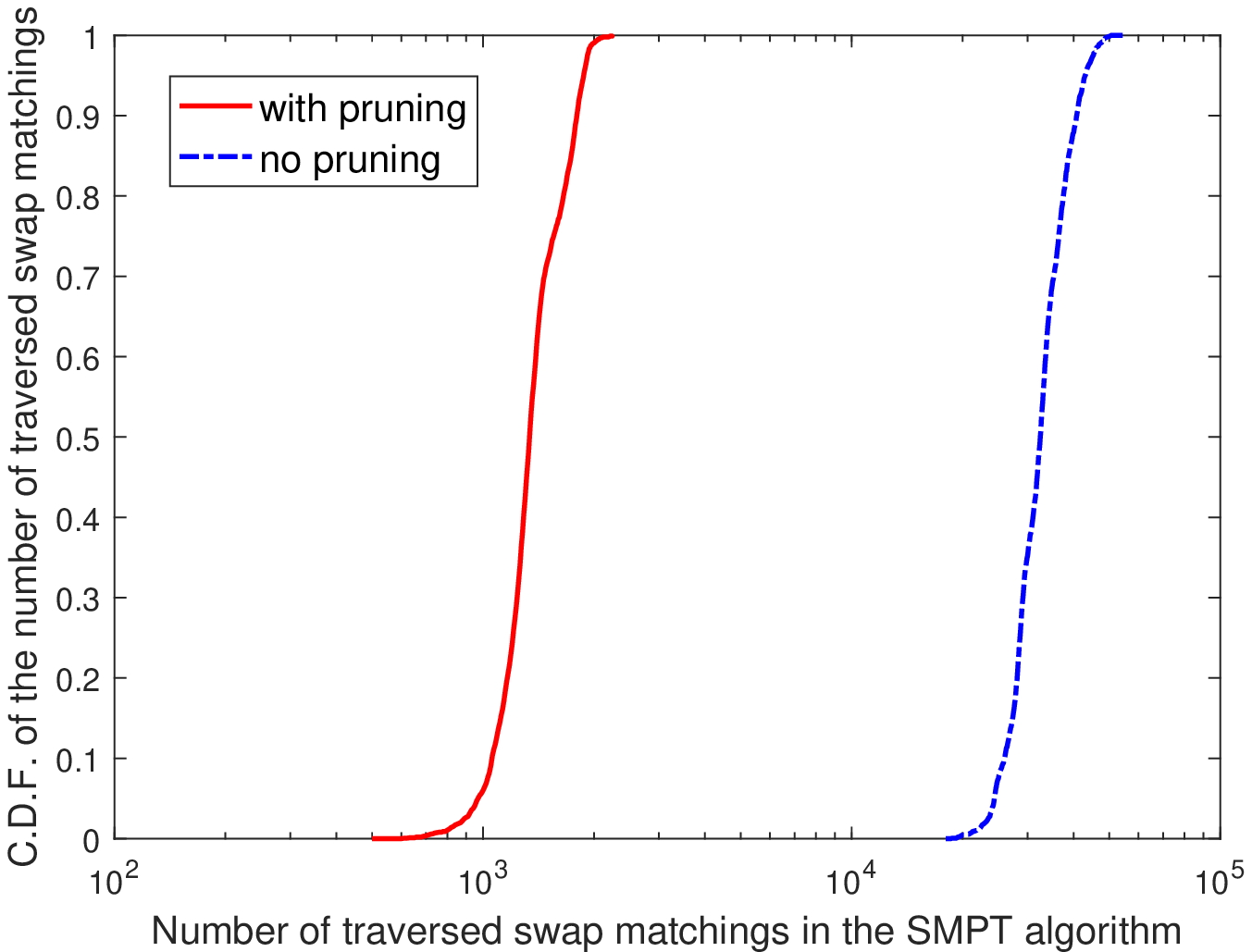}}
\caption{a) C.D.F. of the total number of executed swap operations in the proposed SMPC algorithm; b) C.D.F. of the total number of traversed swap operations in the proposed SMPC algorithm with 18 TSTs} \label{complex} 
\end{figure}

Denote the random variables $\tilde Y_1$ and $\tilde Y_2$ as the total numbers of executed and attempted swap operations in the proposed SMPC algorithm, respectively. As mentioned in Section~\ref{SMPT_complexity}, $\tilde Y_2$ directly reflects the computational complexity and $\tilde Y_1$ is proportional to the complexity.  Fig.~\ref{complexity} shows the cumulative distribution function (C.D.F.) of $\tilde Y_1$, $\Pr \left( {\tilde Y_1 \le \tilde y_1} \right)$, versus $\tilde y_1$ for different numbers of TSTs. Fig.~\ref{complexity1} shows the C.D.F. of $\tilde Y_2$, $\Pr \left( {\tilde Y_2 \le \tilde y_2} \right)$, versus $\tilde y_2$ for the cases where the pruning procedure is performed and not performed, respectively, given 18 TSTs. As the number of TSTs grows, the speed of convergence gets faster, especially when the pruning procedure is performed. This also reflects the low computational complexity of the proposed SMPC algorithm with the effective pruning procedure.

\begin{figure}
\centering
\subfigure[]{
\label{rate_user} 
\includegraphics[width=2.1in]{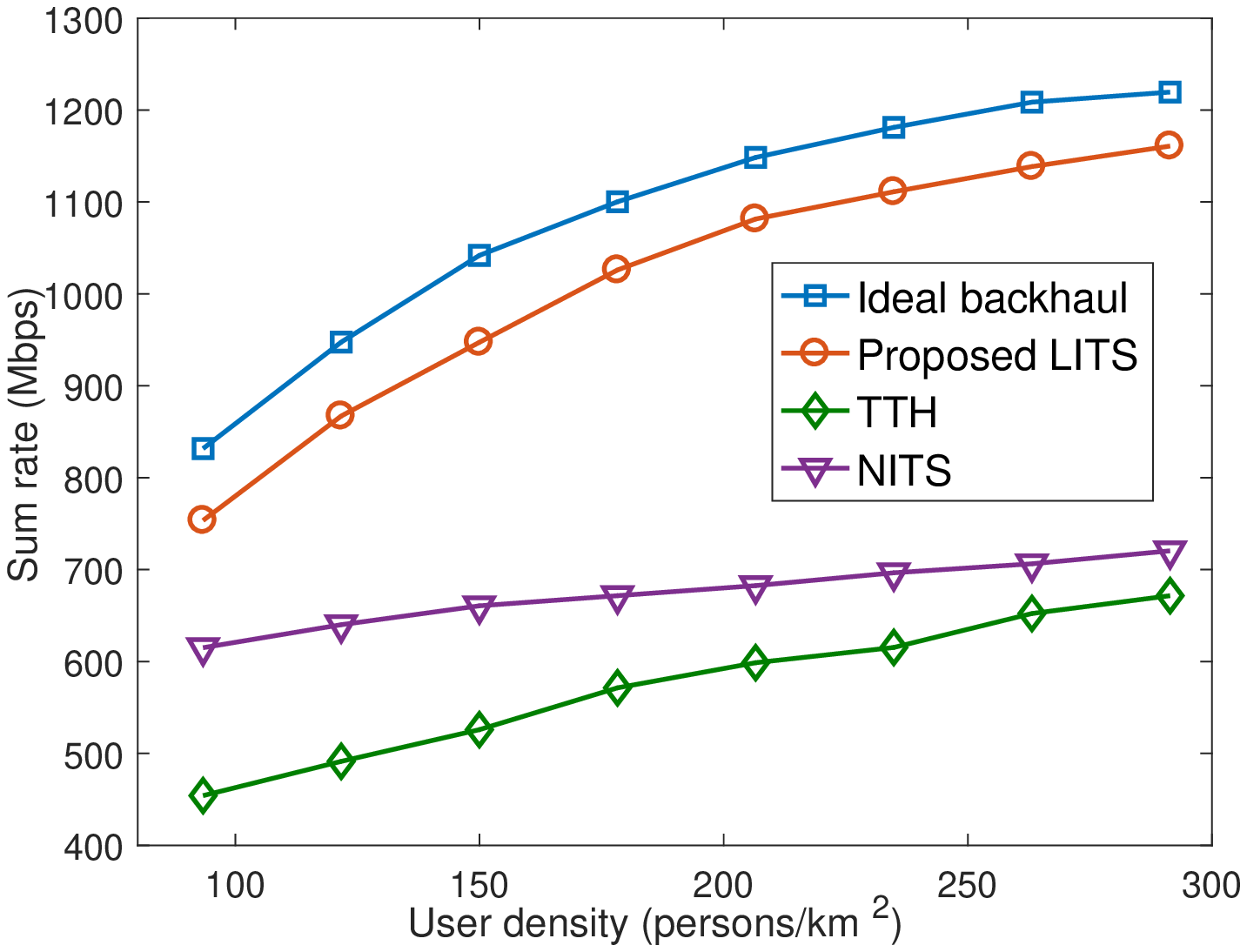}}
\hspace{-0.3in}
\subfigure[]{
\label{capacity_user} 
\includegraphics[width=2.2in]{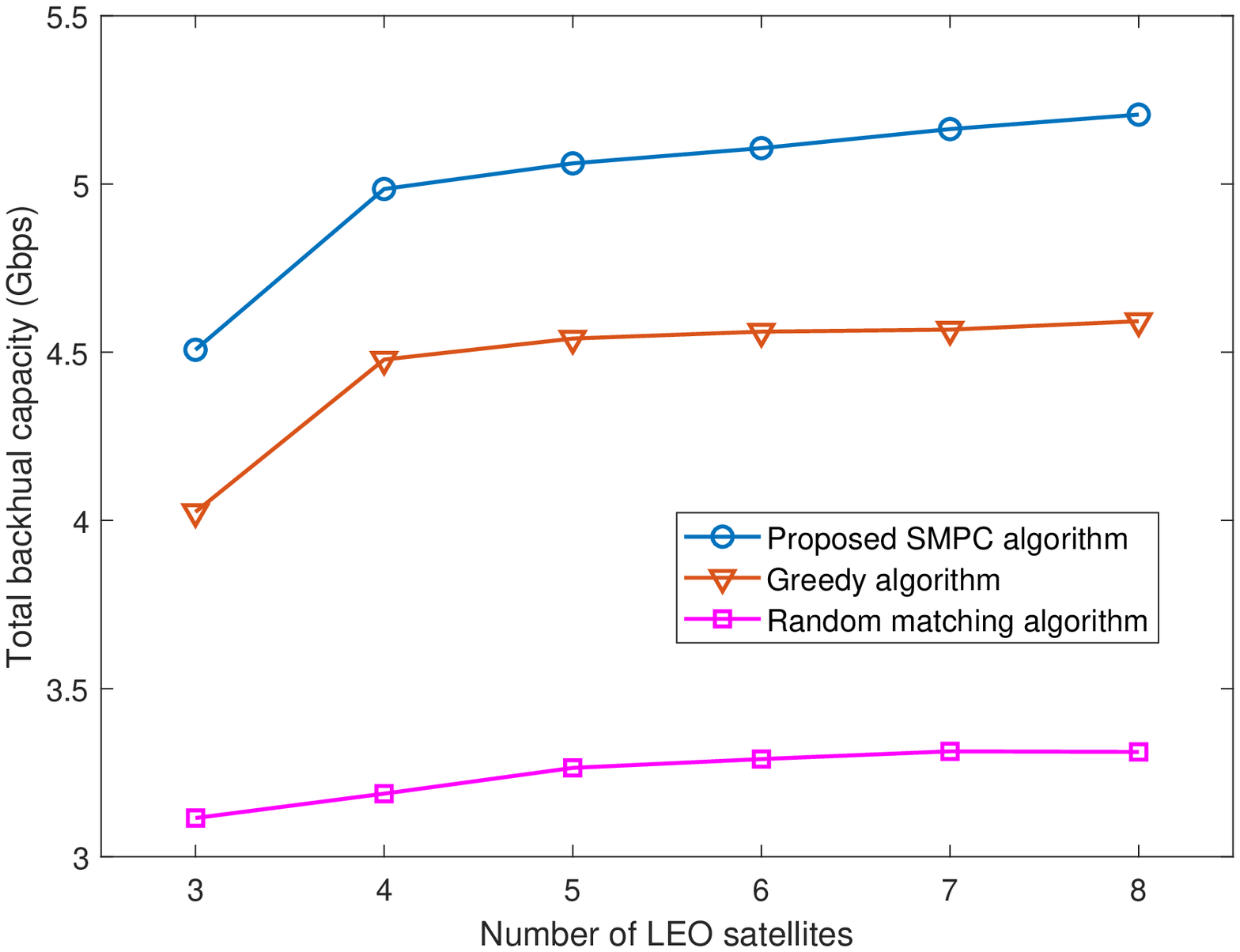}
}
\hspace{-0.4in}
\subfigure[]{
	\label{area_capacity} 
	\includegraphics[width=2.2in]{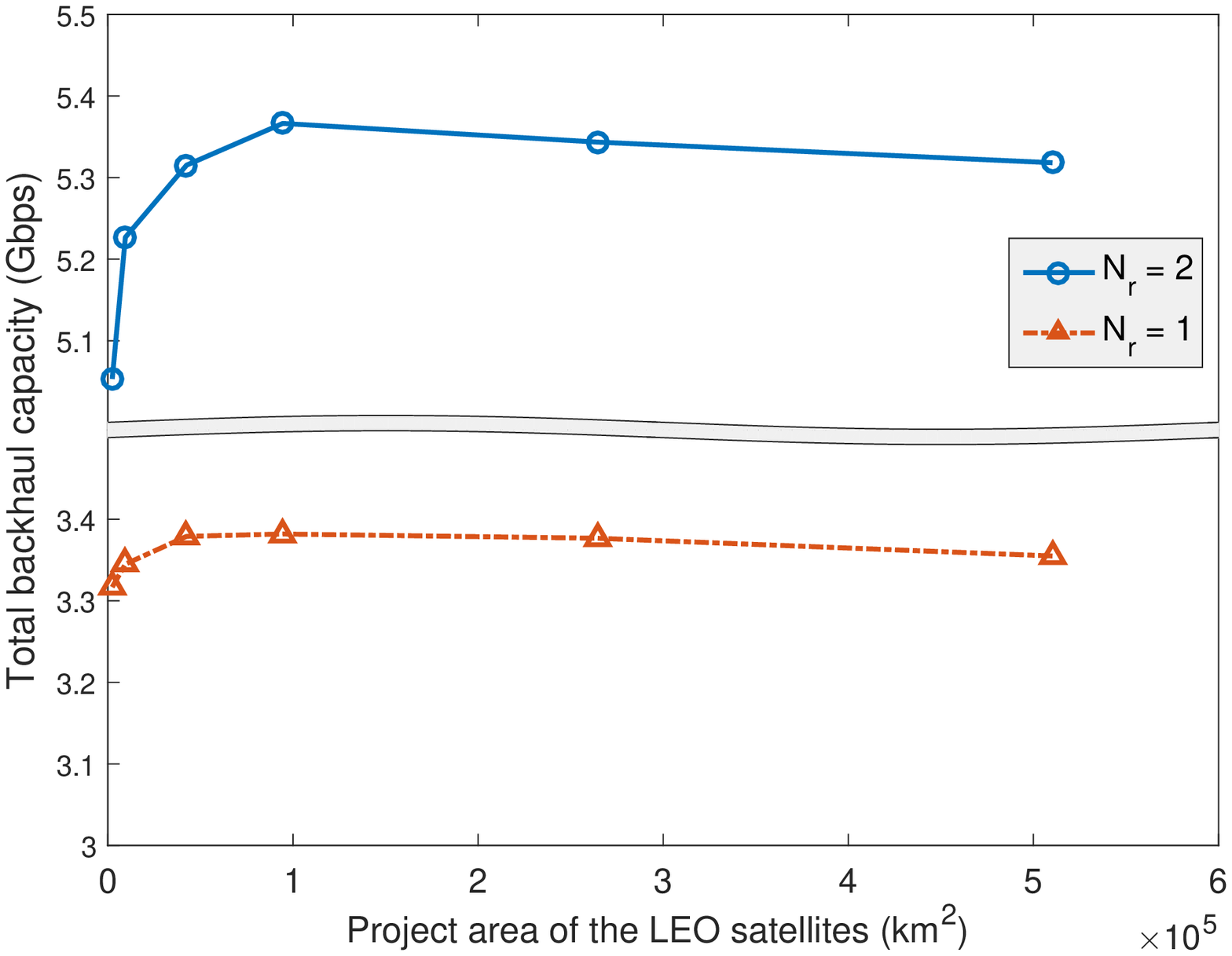}
}
\caption{a) Sum rate of all users v.s. user density; b) Total backhaul capacity v.s. number of LEO satellites; c) Total backhaul capacity v.s. projected area of the LEO satellites} \label{rate_capacity} 
\end{figure}

Fig.~\ref{rate_user} illustrates the sum rate of all users v.s. user density obtained by different schemes. In this paper, the sum rate increases with the user density in the given area. It is observed that the performance of our proposed LITS scheme is close to the case with ideal backhaul and is much better than that of the TTH scheme and the NITS scheme. This implies that the LEO-based backhaul can significantly improve the system performance due to its high backhaul capacity. Besides, via the integration of the TSCs and the LSCs, the user scheduling and frequency resource allocation can be performed more efficiently compared to the non-integrated scheme.

Fig.~\ref{capacity_user} shows the total backhaul capacity v.s. the number of LEO satellites with different numbers of satellites that each TST can access, i.e., $N_r$. From the figure, the total backhaul capacity grows with the number of satellites and turns out to be diminishing returns, indicating the optimal density of satellites shows up at the inflection point. As $N_r$ grows,  the total backhaul capacity increases rapidly owning to the performance gain brought by the multi-connectivity of the TSTs. 

In Fig.~\ref{area_capacity}, We also present how the projected area of a fixed number of LEO satellites influences the total backhaul capacity. As the projected area becomes smaller, the distance between a TST and a satellite is shortened, bringing more high-quality backhaul links. When the projected area is small enough, the inter-satellite interference dominates the backhaul capacity as the angular separation of any two satellites shrinks, leading to declined backhaul capacity. This offers a reference for the LEO constellation deployment for backhaul capacity maximization.

\begin{figure}
	\centering
	\subfigure[]{
		\label{backhaul_delay} 
		\includegraphics[width=3.1in]{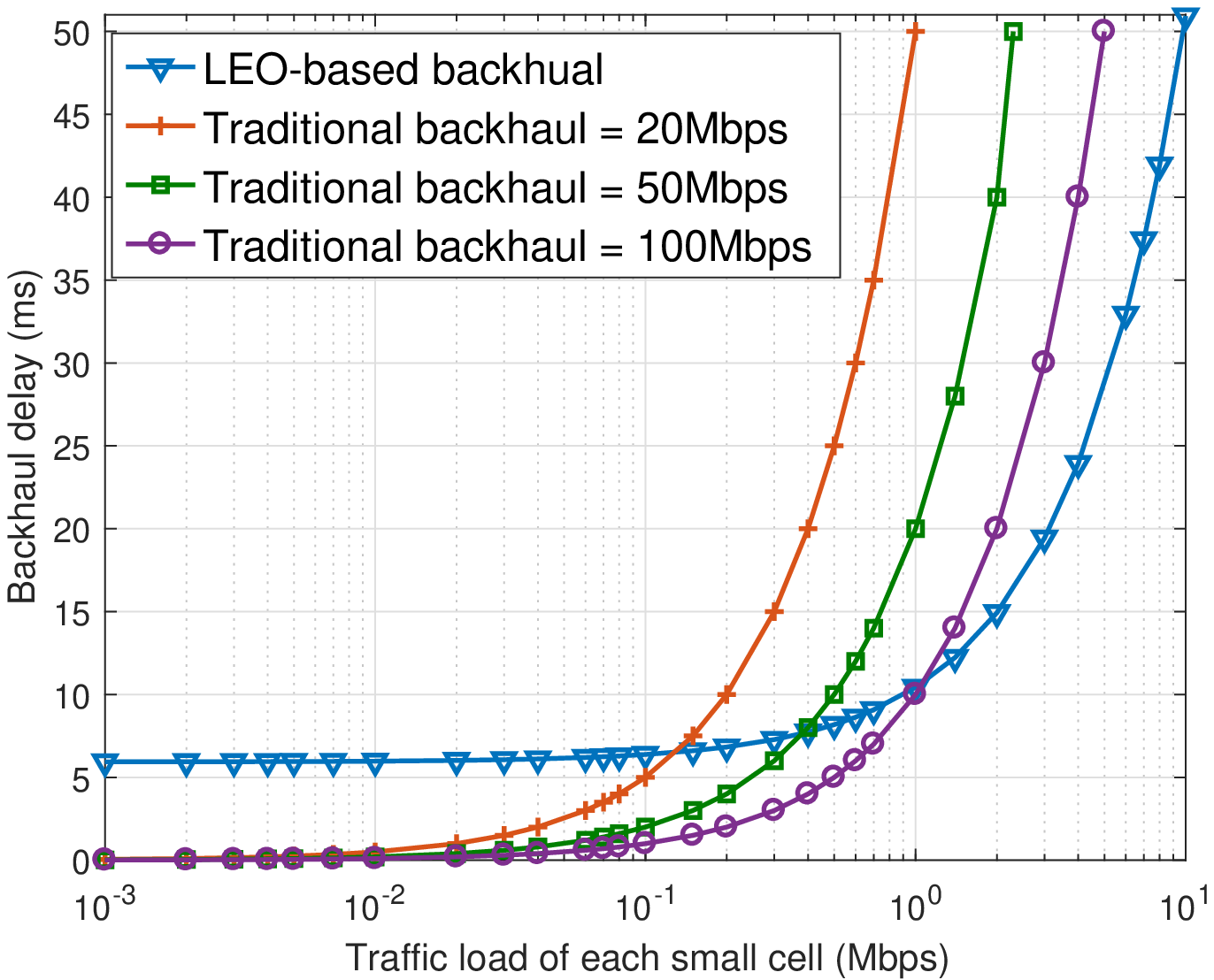}}
	\hspace{-0.2in}
	\subfigure[]{
		\label{LEOpropotion} 
		\includegraphics[width=3.2in]{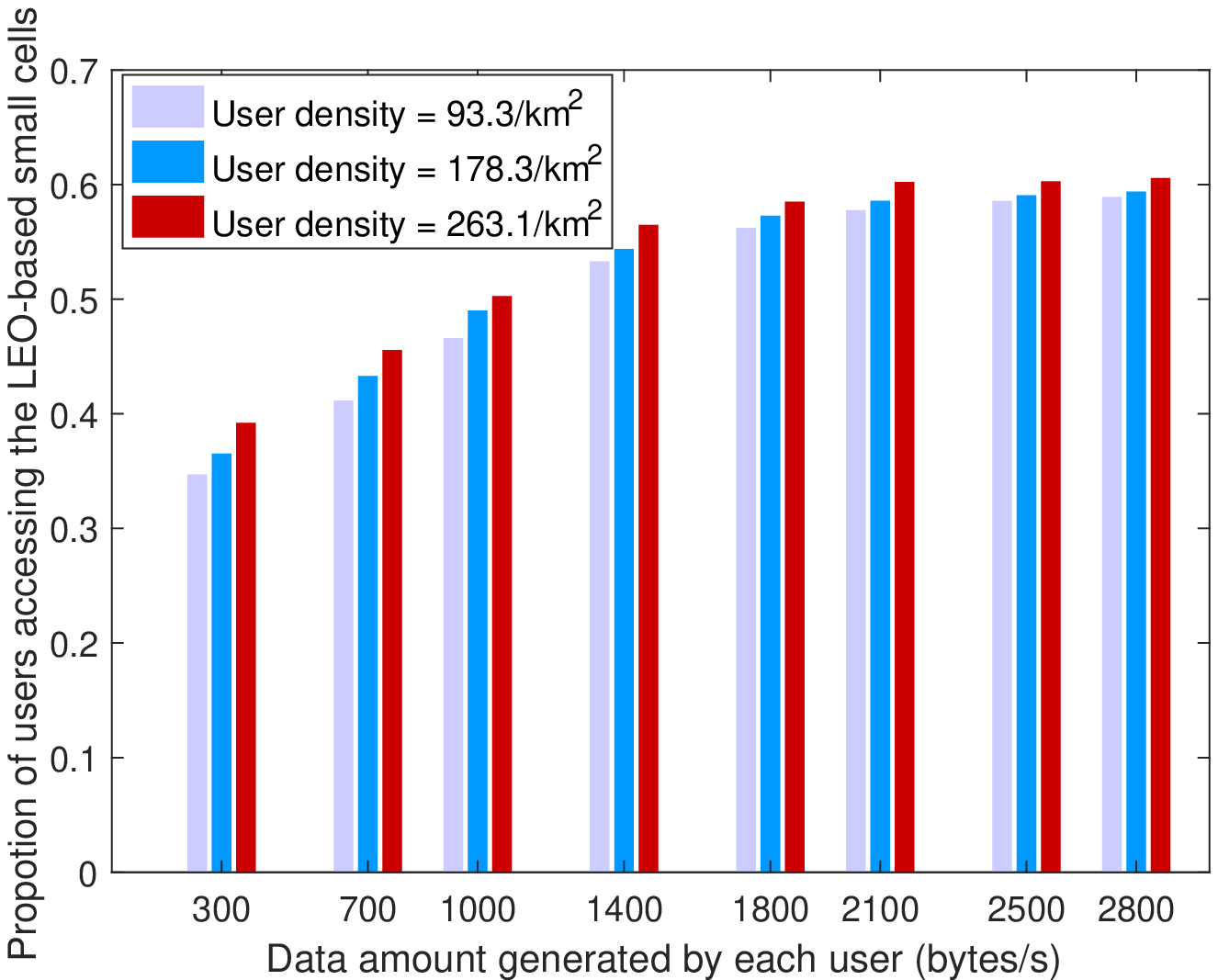}}
	\caption{a) Backhaul delay v.s. traffic load of each small cell; b) Proportion of users accessing the LSCs v.s. amount of data generated by each user} \label{delay} 
\end{figure}

Fig.~\ref{delay} presents how the traffic load influences the system performance in terms of the backhaul delay and the user scheduling strategies. In Fig.~\ref{backhaul_delay}, the backhaul delay is evaluated in both the LSCs and the TSCs with different traffic loads at each TST or SBS. Note that the backhaul capacity of the LSC is usually larger compared to that of the TSCs, leading to low transmission delay. However, the round trip time of the LEO-based small cell is also much larger than that of the TSC due to the long distance between the TST and the LEO satellites. Therefore, when the traffic load is small, the backhaul delay of the TSCs is lower than that of the LSCs. In contrast, when the traffic load is large enough, the advantage of the LEO-based backhaul is reflected since the round trip time can be neglected compared to the transmission delay. This provides a criterion for the network to select the backhaul strategy according to the traffic load.

Fig.~\ref{LEOpropotion} further evaluates how the traffic load influences the user scheduling strategy. From the figure, the proportion of users accessing the LEO-base small cells grows with the amount of data generated by each user. Here we ignore the terrestrial transmission delay since it is too small. When the data generated by each user increases, the traffic load at each TST becomes larger, and thus, the equivalent backhaul capacity for each TST grows according to $\left( \ref{backhaul_capacity} \right)$. Therefore, more users are scheduled to access the LSCs for smaller delay. This is also consistent with Fig.~\ref{backhaul_delay} in the sense that the LEO-based backhaul has a large advantage over the traditional backhaul, especially with high traffic load.
\vspace{-0.2cm}
\section{Conclusion}
In this paper, we have proposed the ultra-dense LEO based integrated terrestrial-satellite network architecture where the LEO-based backhaul provides a new vision to extend the traditional networks. We have formulated an optimization problem to maximize the sum rate and the number of the accessed users subject to the simultaneously optimized backhaul capacity of each LSC. Such an intractable optimization problem has been decomposed into two subproblems connected by the Lagrangian multiplexers. We have converted these two subproblems into matching problems and solved them separately by proposing two different low-complexity matching algorithms.

We have obtained two important conclusions. First, it is encouraged to arrange the users to access the TSC for low propagation delay when the traffic load is small. As the traffic load grows, more users are inclined to access the LSC for lower transmission delay. Second, for a given number of LEO satellites in the visual range of TSTs, there exists an optimal satellite deployment for the total backhaul capacity maximization.

\vspace{-0.2cm}
\begin{appendices}
\section{}
The convergence of the TUASA algorithm depends on the matching process. According to the coverage matrix $\textbf{\emph{A}}$, the number of potential users associated to each BS is limited, and thus, the preference list of each BS-subchannel unit is complete and transparent. In each iteration, the matched BS-subchannel unit proposes to an unmatched user and a BS in its preference list. As the number of iterations grows, the set of remained choices of each BS-subchannel unit becomes smaller. Therefore, the iterations stop when there are no available choices in the preference list for each matched BS-subchannel unit.
%
\vspace{-0.7cm}
\section{}
Suppose that we have $\Psi \left( {{j_1}} \right) = \left( {{m_1},k_1} \right)$ and $\Psi \left( {{j_2}} \right) = \left( {{m_2},k_2} \right)$ in the final matching $\Psi^*$. Consider the matching pairs $\left( {{j_1},\left( {{m_2},k_2} \right)} \right)$ and $\left( {{j_2},\left( {{m_1},k_1} \right)} \right)$.

i) $\rho_2 = 0$: There are three possible reasons why the above two matching pairs do not show up in $\Psi^*$. The first one is that in the initialization phase user $j_1$ is already matched with $\left( m_1, k_1\right) $, and thus, $j_1$ has never been proposed in the matching process. If so, we can infer that either $\left| {h_{{j_1},{m_1},{k_1}}^B} \right| > \left| {h_{{j_2},{m_1},{k_1}}^B} \right|$ (i.e., user $j_1$ and BS $m_1$ are most preferred by subchannel $k_1$) or $\left| {h_{{j_2},{m_2},{k_2}}^B} \right| > \left| {h_{{j_1},{m_2},{k_2}}^B} \right|$. The second possible reason is that the pair $\left( {{j_1},\left( {{m_2},k_2} \right)} \right)$ has never been proposed throughout the algorithm. This implies that user $j_2$ has already been matched before the pair $\left( {{j_1},\left( {{m_2},k_2} \right)} \right)$ is considered by some $\left( m, k\right)$ unit, and thus, we have $\left| {h_{{j_2},{m_2},{k_2}}^B} \right| > \left| {h_{{j_1},{m_2},{k_2}}^B} \right|$. The third reason is that the pair $\left( {{j_1},\left( {{m_2},k_2} \right)} \right)$ was proposed once but rejected by user $j_1$. If so, we can infer that either $\left| {h_{{j_1},{m_1},{k_1}}^B} \right| > \left| {h_{{j_2},{m_1},{k_1}}^B} \right|$ (user $j_2$ is not matched yet) or $\left| {h_{{j_2},{m_2},{k_2}}^B} \right| > \left| {h_{{j_1},{m_2},{k_2}}^B} \right|$ (user $j_2$ has been matched earlier).

For the case where $\left| {h_{{j_1},{m_1},{k_1}}^B} \right| > \left| {h_{{j_2},{m_1},{k_1}}^B} \right|$, since the co-channel interference to BS $m_1$ keeps unchanged for $\left( {{j_1},\left( {{m_1},k_1} \right)} \right)$ and $\left( {{j_2},\left( {{m_1},k_1} \right)} \right)$ as shown in $\left( \ref{single_rate} \right)$, we have $R_{{m_1},{j_1},{k_1}}^B > R_{{m_1},{j_2},{k_1}}^B$. Therefore, condition ii) in Definition 1 is not satisfied, implying that $\Psi^*$ is group stable.

ii) $\rho_2 \ne 0$: when $\rho_1 = 0$, the deduction is very similar to case i). We can always find an existing pair $\left( {{j'},\left( {{m'},k_2} \right)} \right)$ over subchanel $k_2$ such that $\left( m',k_2\right) $ has proposed to $\left( {{j_2},\left( {{m_2},k_2} \right)} \right)$ earlier. This indicates that user $j_1$ will bring larger interference to BS $m'$ than user $j_2$, and thus, $\left( {{j_1},\left( {{m_2},k_2} \right)} \right)$ cannot bring higher utility to $\left( m',k_2\right)$. We then say that $\Psi^*$ is an equilibrium point.

When $\rho_1 \ne 0$, we take the pair $\left( {{j_1},\left( {{m_2},k_2} \right)} \right)$ as an example. Similar to the three cases in i), we can infer that there must exist an existing pair $\left( {j',\left( {m',{k_2}} \right)} \right)$ in $\Psi^*$ satisfying $\left| {h_{{j_2},{m_2},{k_2}}^B} \right|/\left| {h_{{j_2},m',{k_2}}^B} \right| > \left| {h_{{j_1},{m_2},{k_2}}^B} \right|/\left| {h_{{j_2},m',{k_2}}^B} \right|$. Therefore, we have either $\left| {h_{{j_2},{m_2},{k_2}}^B} \right| > \left| {h_{{j_1},{m_2},{k_2}}^B} \right|$ or $\left| {h_{{j_2},m',{k_2}}^B} \right| < \left| {h_{{j_2},m',{k_2}}^B} \right|$. In other words, either condition ii) or i) in Definition 1 is violated, implying that $\Psi^*$ is an equilibrium point.

%

\vspace{-0.5cm}
\section{}
\textbf{Solution to the PC1 problem:} We first divide this problem into two PC3-type problems with the objective functions $R_{{q_2}}^T\left( {p_{{m_1},{n_2},q_2}^T} \right)$ and $R_{{q^*}}^T\left( {p_{{m_1},{n^*},{q^*}}^T} \right)$, respectively. The constraints over $p_{{m_1},{n_2}}^T$ and ${p_{{m_1},{n^*},{q^*}}^T}$ are both set as $\left[ {0,{P_T}} \right]$. Let ${\nabla _{p_{{m_1},{n^*},{q^*}}^T}}R_{{q^*}}^T\left( {p_{{m_1},{n^*},{q^*}}^T} \right) = {\nabla _{p_{{m_1},{n_2},q_2}^T}}R_{{q_2}}^T\left( {p_{{m_1},{n_2},q_2}^T} \right) = 0$ such that the extreme points can be found. We can then obtain the maximum values of the above two PC3-type problems by traversing all extreme points and two boundary points (i.e., 0 and $P_T$). Denote the solutions as ${p_{{m_1},{n_2},q_2}^{T*}}$ and ${p_{{m_1},{n^*},{q^*}}^{T*}}$. If $p_{{m_1},{n^*},{q^*}}^{T*} + p_{{m_1},{n_2},q_2}^{T*} \le p_{{m_1}}^{budget}$, then we naturally obtain the optimal solution of PC1; otherwise, we set $p_{{m_1},{n^*},{q^*}}^T = p_{{m_1}}^{budget} - p_{{m_1},{n_2},q_2}^T$ and substitute it into the original objective function of $\left( \ref{power_problem1}\right)$. The problem then becomes a maximization problem with one continuous variable, which can be easily solved by searching the extreme points.

\textbf{Solution to the PC2 problem:} To save space, we set ${p_1} = p_{{m_1},{n^*},{q^*}}^T$ and ${p_2} = p_{{m_1},{n_2},{q^*}}^T$ in $\left( \ref{PC2_objective} \right)$. The objective function in $\left( \ref{PC2_objective} \right)$ can be reformulated as
\begin{equation}
\begin{split}
&R_{{q^*}}^T\left( {{p_1},{p_2}} \right) = \left[ {\lambda _{{m_1}}}{{\log }_2}\left( {A{p_1} + B{p_2} + {{I'}_{{n^*}}}} \right) + {\lambda _{{m_1}}}{{\log }_2}\left( {C{p_2} + D{p_1} + {{I'}_{{n_2}}}} \right) \right. \\
&\left. + \sum\limits_{m' \ne {m_1}} {{\lambda _{m'}}\sum\limits_{n'} {{{\log }_2}\left( {{F_{m',n'}}{p_1} + {G_{m',n'}}{p_2} + {{I'}_{n'}} + {S_{m',n',{q_2}}}} \right)} }  \right] - \left[ {\lambda _{{m_1}}}{{\log }_2}\left( {B{p_2} + {{I'}_{{n^*}}}} \right) \right.\\
&\left. + {\lambda _{{m_1}}}{{\log }_2}\left( {D{p_1} + {{I'}_{{n_2}}}} \right) + \sum\nolimits_{m' \ne {m_1}} {{\lambda _{m'}}\sum\nolimits_{n'} {{{\log }_2}\left( {{F_{m',n'}}{p_1} + {G_{m',n'}}{p_2} + {{I'}_{n'}}} \right)} }  \right] \\
&= {g_1}\left( \textbf{\emph{p}} \right) - {g_2}\left( \textbf{\emph{p}} \right), 
\end{split}
\end{equation}
where we set the vector $\textbf{\emph{p}} = \left( {{p_1},{p_2}} \right)$ and all $I'$ refers to the sum of interference and noise. Since each item in ${g_1}\left( \textbf{\emph{p}} \right)$ and ${g_2}\left( \textbf{\emph{p}} \right)$ is a concave function of $\textbf{\emph{p}}$, both ${g_1}\left( \textbf{\emph{p}} \right)$ and ${g_2}\left( \textbf{\emph{p}} \right)$ are concave. Therefore, $R_{{q^*}}^T\left( {{p_1},{p_2}} \right)$ can be considered as a difference of convex function. Since the constraint in PC2 represents a convex and close set, a classic difference of convex algorithm~\cite{parida2014power} can be utilized to solve the PC2 problem.
\vspace{-0.6cm}
\section{}
Given two matching pairs in $\Phi^*$, say, $\left( {{m_1},\left( {{n_1},q} \right)} \right)$ and $\left( {{m_2},\left( {{n_2},q} \right)} \right)$ with equal transmit power, we check whether $\left( {{m_2},\left( {{n_1},q} \right)} \right)$ and $\left( {{m_1},\left( {{n_2},q} \right)} \right)$ can form blocking pairs. Since all TSTs share the same coordinates, we use the index $m$ in the antenna coefficients $G$. Suppose that the above two existing pairs are formed in phase 1. According to the preference relation in $\left( \ref{TST_preference_matrix}\right)$, we have either ${\left| {{h_{{m_1},{n_1},q}}} \right|^2} > {\left| {{h_{{m_2},{n_1},q}}} \right|^2}$ or ${\left| {{h_{{m_2},{n_2},q}}} \right|^2} > {\left| {{h_{{m_1},{n_2},q}}} \right|^2}$. We present the rates of satellite $n_1$ before and after blocking as below, i.e., $R_{{n_1},q}^{before} = {\log _2}\left( {1 + \frac{{G_{m,{n_1}}^{m,{n_1}}{{\left| {{h_{{m_1},{n_1},q}}} \right|}^2}}}{{{{I'}_{{m_1},q}} + G_{m,{n_1}}^{m,{n_1}}{{\left| {{h_{{m_2},{n_1},q}}} \right|}^2}}}} \right)$ and $R_{{n_1},q}^{after} = {\log _2}\left( {1 + \frac{{G_{m,{n_1}}^{m,{n_1}}{{\left| {{h_{{m_2},{n_1},q}}} \right|}^2}}}{{{{I'}_{{m_1},q}} + G_{m,{n_1}}^{m,{n_1}}{{\left| {{h_{{m_1},{n_1},q}}} \right|}^2}}}} \right)$. Based on the channel conditions, we have either $R_{{n_1},q}^{before} > R_{{n_1},q}^{after}$ or $R_{{n_2},q}^{before} > R_{{n_2},q}^{after}$. Therefore, $\left( {{m_2},\left( {{n_1},q} \right)} \right)$ and $\left( {{m_1},\left( {{n_2},q} \right)} \right)$ are not blocking pairs since they cannot bring higher utility to all corresponding satellites and TSTs. Moreover, even if the above two existing pairs are obtained from an approved swap matching, the above statement still holds since the swap matching further improves the utility of two matching pairs obtained from phase 1.

\vspace{-0.6cm}
\section{}
Based on the size of three sets ${\cal{M}'}$, $\cal{N}$, and $\cal{Q}$, it is easy to obtain the number of type-1 (or 2) swap matchings as $NQ\left( {M - M'} \right)$. For type-3 swap matchings, each TST with $N_r$ links can construct ${N_r}\left( {{N_r} - 1} \right)/2$ swap matchings, and thus, the total number of swap matchings is ${N_r}\left( {{N_r} - 1} \right)\left( {M - M'} \right)/2$. We now consider the type-4 and type-5 swap matchings. For each TST $m$ with $N_r$ links, there exist at most $\max \left\{ {NQ - {N_r},\left( {M - M' - 1} \right){N_r}} \right\}$ links not including $m$, each of which can form a swap matching with one of TST $m$'s links. Therefore, the total number of swap matchings is ${N_r} \cdot \sum\nolimits_{i = 1}^{M - M' - 1} {\max \left\{ {NQ - i{N_r},\left( {M - M' - i} \right){N_r}} \right\}}$ = ${N_r}\left( {M - M' - 1} \right)/2 \cdot \max \left\{ {2NQ - \left( {M - M'} \right){N_r},\left( {M - M'} \right){N_r}} \right\}$.
\end{appendices}

\vspace{-0.4cm}

\bibliographystyle{ieeetr}%
\bibliography{bibilio}

\end{document}